\documentclass[a4paper,11pt]{article}
\pdfoutput=1
\usepackage{jcappub}
\usepackage{epsfig}
\usepackage{graphicx}
\usepackage{color}
\usepackage{amssymb}
\usepackage{amsmath}
\usepackage{mathtools}
\usepackage{hyperref}
\usepackage{url}
\usepackage{natbib}
\usepackage[normalem]{ulem}

\newcommand{\bfe}{{\mathbf e}}

\newcommand{\bn}{{\mathbf n}}

\newcommand{\HH}{{\cal H}}

\newcommand{\de}{\delta}
\newcommand{\De}{\Delta}

\newcommand{\ga}{\gamma}

\newcommand{\ka}{\kappa}

\newcommand{\La}{\Lambda}

\newcommand{\Om}{\Omega}
\newcommand{\om}{\omega}

\newcommand{\be}{\begin{equation}}
\newcommand{\ee}{\end{equation}}

\newcommand{\bea}{\begin{eqnarray}}
\newcommand{\eea}{\end{eqnarray}}
\newcommand{\bean}{\begin{eqnarray*}}
\newcommand{\eean}{\end{eqnarray*}}
\newcommand{\dd}{\partial}

\newcommand*\bra[1]{\left(#1 \right)}

\newcommand\spart{\;\raise1.0pt\hbox{$/$}\hskip-6pt\partial}

\title{Measuring the lensing potential with tomographic galaxy number counts}

\author[]{Francesco Montanari,}
\author[]{Ruth Durrer}

\affiliation[]{D\'epartement de Physique Th\'eorique and Center for Astroparticle Physics,
Universit\'e de Gen\`eve\\ 24 quai Ernest  Ansermet, 1211 Gen\`eve 4, Switzerland}

\emailAdd{francesco.montanari@unige.ch}
\emailAdd{ruth.durrer@unige.ch}

\abstract{
We investigate how the lensing potential can be measured tomographically with future galaxy surveys using their number counts.
Such a measurement is an independent test of the standard $\Lambda$CDM framework and can be used to discern modified theories of gravity.
We perform a Fisher matrix forecast based on galaxy angular-redshift power spectra, assuming specifications consistent with future photometric Euclid-like surveys and spectroscopic SKA-like surveys.
For the Euclid-like survey we derive a fitting formula for the magnification bias.
Our analysis suggests that the cross correlation between different redshift bins is very sensitive to the lensing potential such that the survey can measure the amplitude of the lensing potential  at the same level of precision as other standard $\Lambda$CDM cosmological parameters.
}

\arxivnumber{}

\begin{document}
\maketitle
\flushbottom

\section{Introduction}
Cosmology has become a data driven science. After the amazing success story of the cosmic microwave background (CMB), see~\cite{Ade:2013ktc,Planck:2015xua,Ade:2015tva,Durrer:2015lza,Durrer:2008aa}, we now also want to profit in an optimal way from present and future galaxy catalogs. Contrary to the CMB which comes from the two dimensional surface of last scattering, galaxy catalogs are three dimensional and therefore contain potentially more, richer information.

In the large galaxy surveys planned at present one distinguishes spectroscopic surveys, which determine the redshift of galaxies very precisely and are used to determine the large scale structure (LSS), i.e., the clustering properties of galaxies, and  photometric surveys which determine the redshift with less precision but which are optimized  for galaxy shape measurements. From shape measurements one can then statistically infer the shear $\ga$ and from it the lensing potential.
In this paper we study to which extent the lensing potential can be obtained by simply looking at the angular correlation function of galaxies at different redshift without invoking shape measurements which are plagued by  difficult systematics and intrinsic alignment~\cite{Schneider:2005ka}.

The lensing potential is especially interesting since its relation to the matter density is determined by Einstein's equations. The lensing potential $\psi$ is given by (see, e.g., \cite{Durrer:2008aa})
\be\label{eq:lenpot}
\psi(\bn,z) = -\int_0^{r(z)}\!\!d\tilde r\frac{r(z)-\tilde r }{r(z)\tilde r}(\Phi+\Psi)(\tilde r\bn,\tau_0-\tilde r)\,
\ee
where $r(z)$ is the comoving distance of the source and $\tau_0$ is the present conformal time.
We neglect spatial curvature, $K=0$, and we define the lensing potential as a function of redshift $z$ and observer direction $\bn$. The variables $\Phi$ and $\Psi$ are the well known Bardeen potentials~\cite{Bardeen:1980kt,Durrer:2008aa}.
By measuring both, the density fluctuation $\de(\bn,z)$ and the lensing potential
we can in principle test General Relativity on cosmological scales.

When observing galaxies we measure their redshift and angular position.  We can then determine the number of galaxies per solid angle and per redshift bin.
The correlation function of these number counts within linear perturbation theory has been determined in Refs.~\cite{Bonvin:2011bg,Challinor:2011bk}. Apart from the galaxy number density and velocity it also depends on the convergence
\be\label{e:kappa}
\ka = -\frac{1}{2}\Delta_\Om\psi \,,
\ee
 which affects the
volume corresponding to a given observed solid angle and redshift bin.
Here $ \Delta_\Om$ is the angular Laplacian.

Number counts have also been studied in \cite{Yoo:2009au,Jeong:2011as,Schmidt:2012ne,Bertacca2012tp} and relativistic expressions to second order have been derived in~\cite{Bertacca:2014dra,Bertacca:2014wga,Bertacca:2014hwa,Yoo:2014sfa,DiDio:2014lka}.
Their potential for cosmological parameter constraints has been analyzed
in several papers \cite{DiDio:2013sea,Raccanelli:2013dza,Raccanelli:2013gja,Yoo:2012se,Yoo:2013tc,Yoo:2013zga,Raccanelli:2015vla,Alonso:2015uua}. In Ref~\cite{DiDio:2013bqa} a code ({\sc Class}gal) for fast computation is presented and  forecasts for DES and Euclid-like catalogs are compared with the traditional LSS analysis.  In Ref.~\cite{Camera:2014bwa,Camera:2014sba} the capacity of SKA (the square kilometer array~\cite{Maartens:2015mra}) to determine primordial non-Gaussianity is analyzed using number counts. Furthermore, relativistic contributions to the number counts can be isolated using special observational techniques~\cite{Bonvin:2013ogt,Bonvin:2014owa,Camera:2015yqa}.
In most of this paper we are concerned with the well known convergence term $\ka$ which strictly speaking is also a ``relativistic contribution''. However, due to the Laplacian, on small and intermediate scales, this term is much larger than the sub-leading velocity terms and the new ``relativistic terms'' derived in~\cite{Yoo:2009au,Bonvin:2011bg,Challinor:2011bk}. The amplitude of the latter is of the order of the gravitational potential, we therefore call them ``potential terms''. We shall see that they are relevant only at very large angular scales and we shall neglect them for most of this work.
Analyses of relativistic number counts in the context of modified gravity can be found in \cite{Lombriser:2013aj,Baker:2015bva}.

Cosmic magnification $\mu=\left[(1-\kappa)^2-|\gamma|^2\right]^{-1}\simeq 1+2\ka$ \cite{Broadhurst:1994qu,Moessner:1997qs,Schneider:2005ka},  has been detected by correlating background quasars and foreground galaxies, e.g., with the SDSS survey \cite{Scranton:2005ci} or other catalogs \cite{Bartelmann:1993pj,Norman:1999hv,Blake:2006qu,Menard:2009yb}.
Recent observations include, e.g., a detection of a redshift-depth enhancement of background galaxies magnified by foreground clusters using BOSS-Survey galaxies \cite{Coupon:2013uwa}, and a measurement of the effects of lensing magnification on the detected number counts of background luminous red galaxies (LRGs) by foreground LRGs and clusters by \cite{Bauer:2013tea} for  the MegaZ (SDSS DR7) catalog.

The analysis presented in this paper goes beyond this work. We constrain the lensing potential using the full tomographic redshift information of the lensing convergence, which allows us to probe the 3D information of a galaxy catalog. In certain situations the lensing term actually dominates the galaxy number count fluctuations and future surveys can be used to determine it. Our method is also very complementary to the usual determination of the lensing potential via shear measurements.  It has  different systematics and measures a somewhat different observable.

This paper is structured as follows:
in the next section we study number counts and determine the situations in which the lensing term dominates.
In section~\ref{s:fm} we present a Fisher matrix study for a photometric Euclid-like and a spectroscopic SKA-like survey.
In section~\ref{s:con} we conclude.
Appendix~\ref{sec:surveys} shows the survey specifications that we assume.
In appendix~\ref{sec:s_bias} we compute the magnification bias for Euclid.
In appendix~\ref{A:spin} we present some properties of spin weighted spherical harmonics used in the main text and appendix~\ref{sec:fisher} outlines our Fisher matrix formalism.

\section{Determining the lensing potential with number counts}
In this section we recall the full expression determining galaxy number counts. We then analyze the different contributions to the power spectra and show how we can isolate the correlation of the density fluctuation with the lensing term. We also discuss the physical importance of a measurement of lensing convergence.

\subsection{Galaxy number counts}
Number counts of a given species of objects, e.g. galaxies is given by $n(z,\bn)=\bar{n}(z)[1+\De(z,\bn)]$ where $\bar{n}(z)$ is the mean galaxy density per redshift and per steradian at redshift $z$ and~\cite{Bonvin:2011bg,Challinor:2011bk,DiDio:2013sea,Camera:2014bwa}
\bea
\De(\bn,z,m_{\rm lim}) &=& b(z)D +\frac{1}{\HH}
\left[\dot\Phi+\dd^2_rV\right]  +(2-5s)\left[\int_0^{r}\hspace{-0.3mm}\frac{d\tilde r}{r} (\Phi+\Psi) - ~ \ka\right] +(f_{\rm evo}-3)\HH V  +   \nonumber \\  &&
(5s-2)\Phi + \Psi+ \left(\frac{{\dot\HH}}{\HH^2}+\frac{2-5s}{r\HH} +5s-f_{\rm evo}\right)\left(\Psi+\dd_rV+
 \int_0^{r}\hspace{-0.3mm}d\tilde r(\dot\Phi+\dot\Psi)\right)  \,.
   \nonumber \\  &&
  \label{e:DezNF}
\eea
A dot denotes a derivative w.r.t. conformal time, $\HH$ is the conformal Hubble parameter, $V$ is the velocity potential for the peculiar velocity in longitudinal gauge, $v_i=-\dd_iV$, and $D$ is the matter density fluctuation in comoving gauge while $b(z)$ denotes the galaxy bias.  More details are found in~\cite{DiDio:2013sea} and~\cite{DiDio:2013bqa}.

Denoting the limiting luminosity and magnitude of the survey by $L_{\rm lim}$ and $m_{\rm lim}$ respectively, we have introduced the evolution bias, which captures the fact that new galaxies form and galaxies merge as the universe expands, hence their number density evolves not simply like $(1+z)^{3}$,
\be
\label{eq:fevo}
f_{\rm evo}(z,L_{\rm lim}) \equiv \frac{\partial\ln\bra{a^3 \bar n(z,L>L_{\rm lim})}}{ \partial \ln a} \, ,
\ee
where $\bar n(z,L> L_{\rm lim})$ indicates the number density (per redshift and per steradian) of galaxies with luminosity above $L_{\rm lim}$ and $a$ is the cosmic scale factor.
We also consider the magnification bias which  takes into account that due to magnification less luminous galaxies still make it into our survey if they are in a region of high magnification and vice versa,
\be
s(z,m_{\rm lim}) \equiv \left.\frac{\partial\log_{10}\bar n(z,L> L_{\rm lim})}{\partial m}\right|_{m_{\rm lim}} \,,
\label{e:s_mlim}
\ee
see appendix~\ref{sec:s_bias} for more details.

The functions $f_{\rm evo}(z,L_{\rm lim})$ and $s(z,m_{\rm lim})$ have to be determined from the catalog specifications, either directly from observations or using simulations. As an example, in~\cite{Camera:2014bwa} the determinations of $b(z,L_{\rm lim})$, $f_{\rm evo}(z,L_{\rm lim}) \equiv b_e(z,L_{\rm lim}) $ and $5s(z,m_{\rm lim})\equiv 2Q(z,L_{\rm lim})$ from simulations for the SKA survey of the 21cm line are discussed.
Our modeling of these functions is described in appendices~\ref{sec:surveys} and~\ref{sec:s_bias}.

The second term in the second square bracket of eq.~(\ref{e:DezNF}) is the lensing term,
\bea
\De^L(\bn,z) &=& -(2-5s)\ka \,.\label{e:DeL}
\eea

Note that this term is specific to number counts. It comes from the change of the transverse surface area which has to be taken into account when relating number counts to density fluctuations. Due to the reciprocity relation this term cancels (at linear order in perturbations theory) in intensities like the 21cm intensity mapping~\cite{Hall:2012wd} or the CMB anisotropies~\cite{Durrer:2008aa}.

We can now expand $\De(\bn,z,m_{\rm lim})$ in spherical harmonics and compute its power spectrum. Suppressing the limiting magnitude (or luminosity) $m_{\rm lim}$ we have
\bea
\De(\bn,z) &=& \sum_{\ell,m}a_{\ell m}(z)Y_{\ell m}(\bn)  \\
  \langle a_{\ell m}(z)a^*_{\ell' m'}(z') \rangle &\equiv& \de_{\ell \ell'}\de_{mm'} C_\ell(z,z')\,,
\eea
the Kronecker-deltas are, as usual, a consequence of statistical isotropy.

In a true catalog we have to use redshift bins of finite thickness. We denote a normalized window function with width $\De z$ centered around the redshift $z_i$ by $W_{\De z}(z,z_i)$ and introduce the correlation power spectra
\be
C_\ell^{ij} \equiv C_\ell(i,j) = \int dzdz' W_{\De z}(z,z_i)W_{\De z}(z',z_j)C_\ell(z,z')\;,
\label{e:Cls}
\ee
which has been thoroughly studied in literature \cite{Asorey:2012rd,DiDio:2013sea,Nicola:2014bma}.
In these observables, the different terms of eq.~(\ref{e:DezNF}) contribute differently.
The first term,  the standard density term  $\De^{\rm D}(\bn,z)=bD$, usually dominates. The standard redshift space distortion term in the Kaiser approximation
\be
\De^{\rm rsd}(\bn,z)=\HH^{-1}\dd_r^2V \;,
\ee
is also very significant, especially for narrow window functions, $\De z\lesssim0.01(1+z)$.
We stress that the usual Kaiser approximation for the power spectrum in Fourier space $P(k)$ or the correlation function in configuration space also assumes that velocities of each galaxy pair are parallel, which implies $\bn_i=\bn_j$ for each pair $i,j$, hence vanishing angular separation.
Such a ``flat sky approximation'' is not necessary for angular correlations, thus including $\De^{\rm rsd}(\bn,z)$ in eq.~(\ref{e:Cls}) we also take into account so-called wide-angle effects \cite{Papai:2008bd,Raccanelli:2010hk,Montanari:2011nz,Yoo:2013tc,Yoo:2013zga}.
We denote all the other velocity terms as ``Doppler terms'', $\De^V(\bn,z)$ and all remaining terms apart from the lensing contribution as ``potential terms'', $\De^P(\bn,z)$, so that
\bea
\De &=& \De^D + \De^{\rm rsd} + \De^L + \De^V + \De^P  \qquad \mbox{with}\\
\De^V(\bn,z) &=& (f_{\rm evo}-3)\HH V + \left(\frac{{\dot\HH}}{\HH^2}+\frac{2-5s}{r\HH}
+5s-f_{\rm evo}\right)\dd_rV \\
\De^P(\bn,z) &=& \HH^{-1}\dot\Phi +\frac{2-5s}{r}\int_0^{r}\hspace{-0.3mm}d\tilde r(\Phi+\Psi) +(5s-2)\Phi + \Psi + \nonumber \\  &&
+ \left(\frac{{\dot\HH}}{\HH^2}+\frac{2-5s}{r\HH} +5s-f_{\rm evo}\right)\left(\Psi+
 \int_0^{r}\hspace{-0.3mm}d\tilde r(\dot\Phi+\dot\Psi)\right) \,.
\label{eq:nc_pot}
\eea
The integrals in the first  and second line of eq.~(\ref{eq:nc_pot}) correspond to the Shapiro time-delay and integrated Sachs-Wolf effects, respectively.

We compute the power spectra in eq.(\ref{e:Cls}) using the {\sc Class} code\footnote{\url{http://class-code.net/}} \cite{Blas:2011rf}.
We generalize version 2.3, already including the {\sc Class}gal modifications \cite{DiDio:2013bqa}, to take into account  the redshift dependence of the bias parameters (appendix~\ref{sec:surveys}).
Furthermore, even if the expressions given above are valid at linear order in perturbation theory, we compute approximate non-linear spectra obtained using Halofit \cite{Halofit} as described in appendix~\ref{sec:fisher}.

\subsection{The contributions to the number counts}
In the rest of this section, except when mentioned explicitly, we take as reference case the Euclid photometric specifications for 10 bins containing equal numbers of galaxies as described in appendix~\ref{sec:surveys}.

\begin{figure}[t!]
\begin{center}
\includegraphics[width=.75\textwidth]{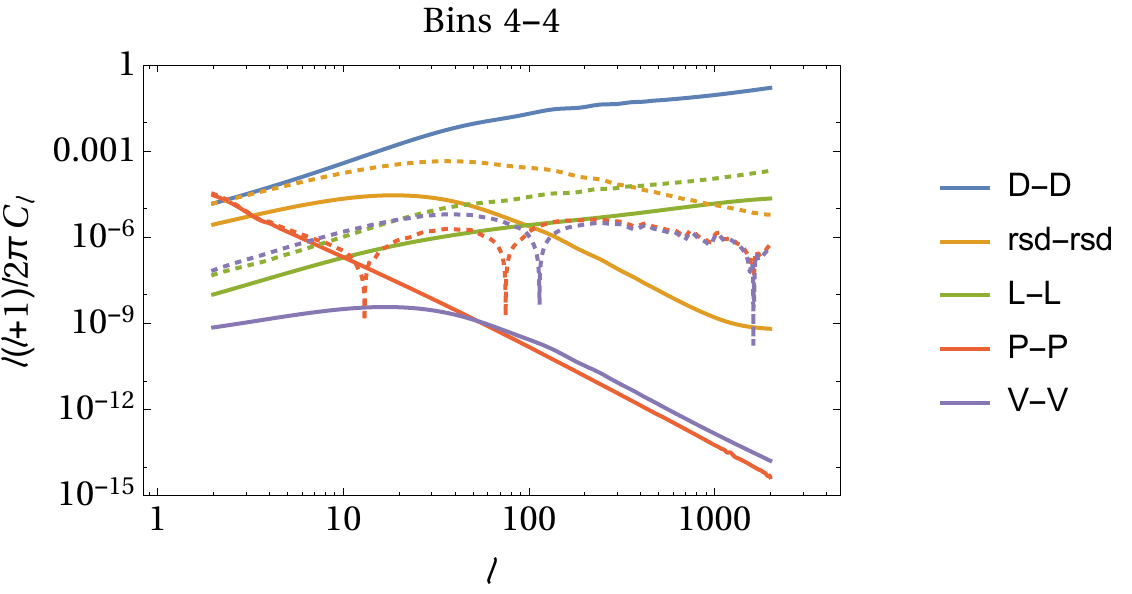}
\end{center}
\caption{Auto-correlations of the $4^{\rm th}$ bin (see figure~\ref{fig:dNdz} for an illustration of our binning).
Due to the large width of the bins, redshift-space distortions are only important at large scales. On the other hand, lensing is the second important effect. Auto-correlations are dominated by the density term.
Dashed lines show the sum of the auto-correlation of a given term plus its correlation with the density (we plot absolute values).
}
\label{f:spec-diag}
\end{figure}

In figure~\ref{f:spec-diag} we show the diagonal spectra of the pure density term,
$C^D_\ell(i,i)$,
the redshift space distortion term,  $C^{\rm rsd}_\ell(i,i)$, the lensing term,  $C^L_\ell(i,i)$, the Doppler term $C^V_\ell(i,i)$ and the potential term $C^P_\ell(i,i)$.
Clearly, the density term dominates, followed by the redshift space distortion and, on smaller scales, by the lensing term. Doppler and potential terms are significantly smaller. This figure is useful to compare the relative importance of different effects, but for the analysis of the full signal also cross correlations between the various terms must be considered.
Since density is the dominant effect, in figure~\ref{f:spec-diag} we plot as dashed lines the spectrum of the correlation of a given term $X$ plus its cross-correlation with the density, schematically $|\langle X(z_i)X(z_j) \rangle+\langle X(z_i)\delta(z_j) \rangle + \langle \delta(z_i)X(z_j) \rangle|$.
As we shall see below, the lensing term is always dominated by the density-lensing cross correlation rather than the lensing-lensing correlation itself.
We have verified that in the case of the spectroscopic SKA surveys, thanks to the good redshift determination, galaxies are not spread over Gaussian bins and the relative importance of local terms (``D'', ``rsd'', ``V'') increases compared to integrated effects (``L'',``P''). See also~\cite{DiDio:2013sea}  for a comparison between photometric and spectroscopic surveys.

In our Fisher matrix analysis in Section~\ref{s:fm} we neglect the potential terms since the likelihood is strongly dominated by $\ell>20$ where they do not contribute appreciably.

\begin{figure}[t!]
\begin{center}
\includegraphics[width=.45\textwidth]{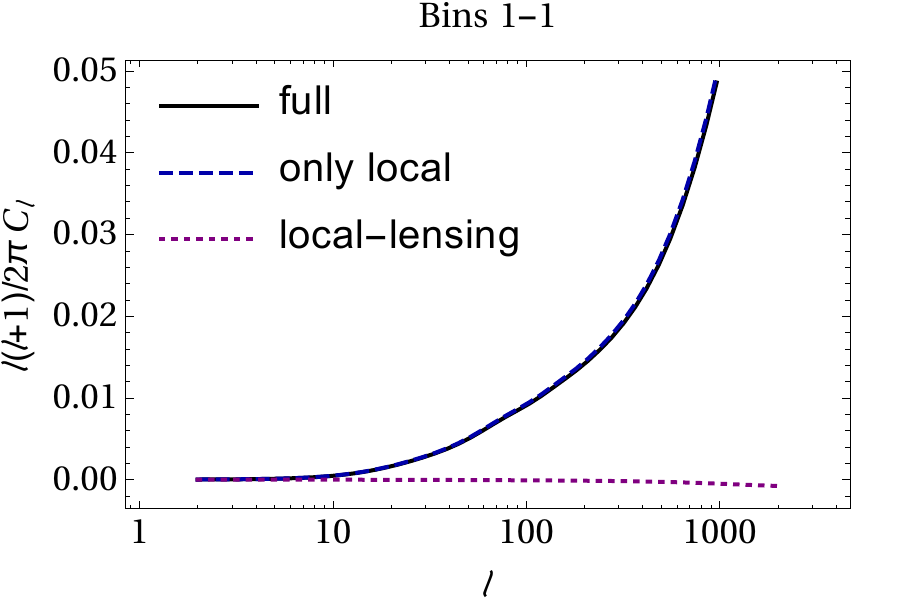}
\includegraphics[width=.45\textwidth]{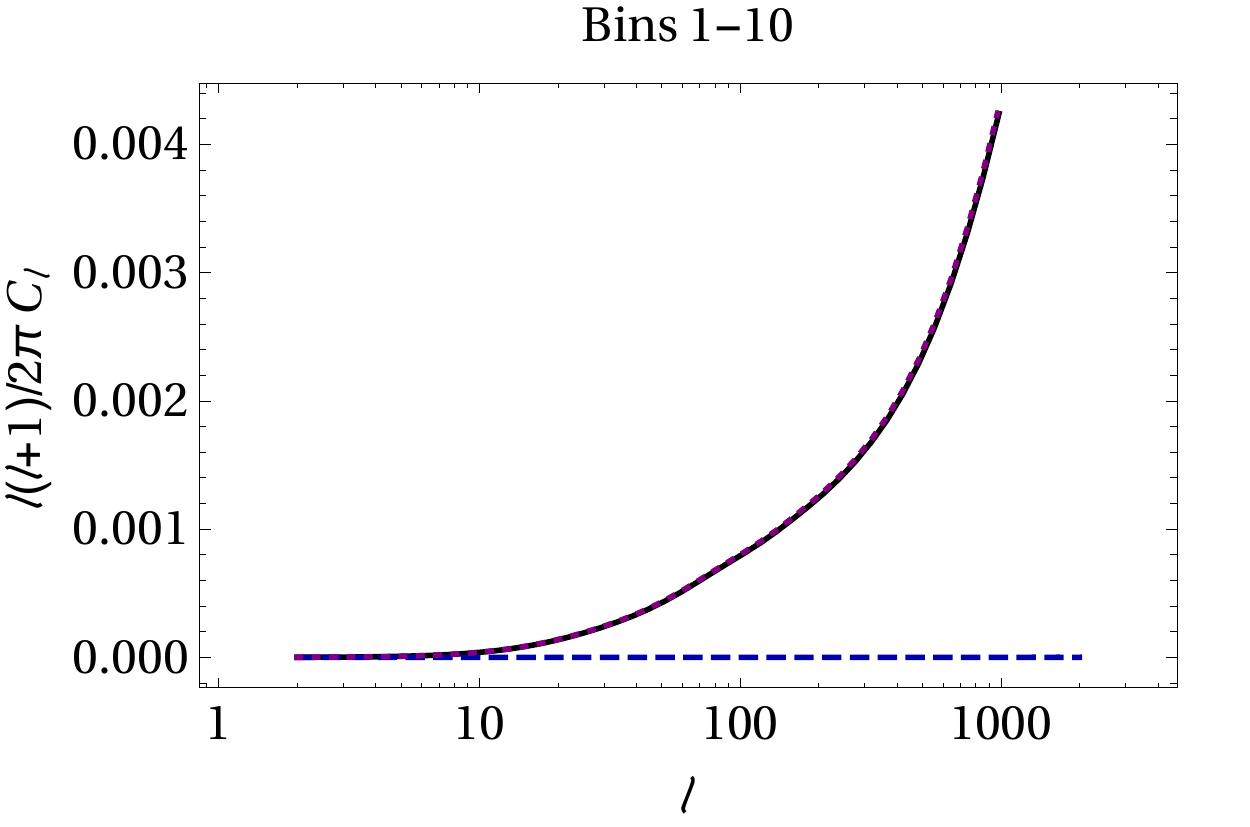}\\
\includegraphics[width=.45\textwidth]{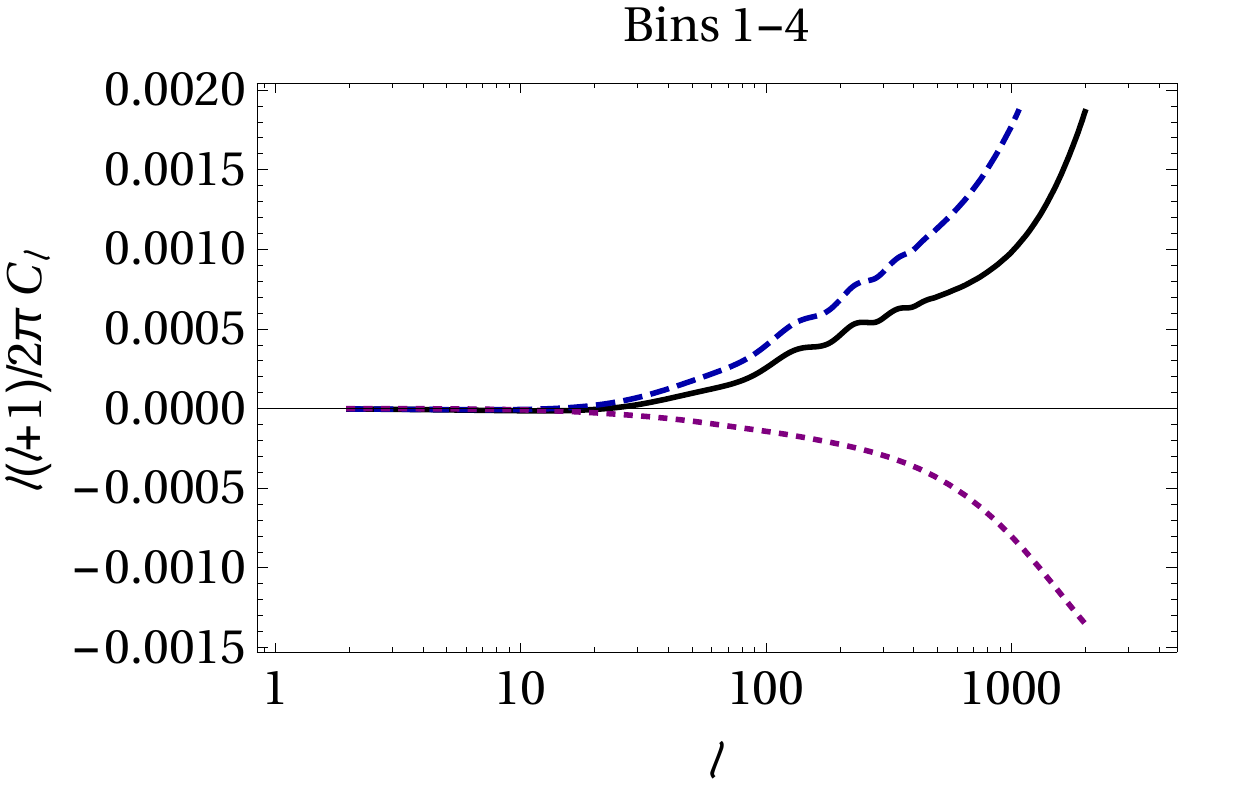}
\includegraphics[width=.45\textwidth]{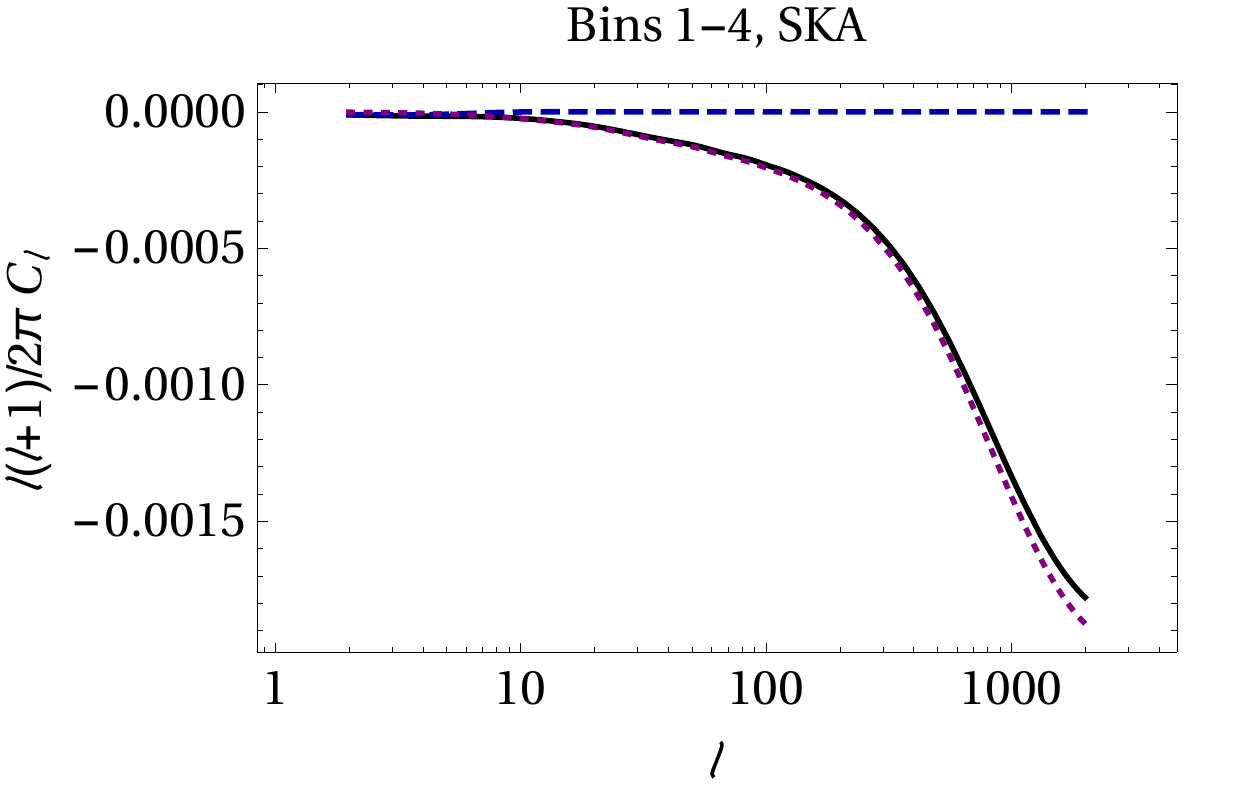}
\end{center}
\caption{{\em Top left panel:} the auto-correlation of the $1^{\rm th}$ bin (see figure~\ref{fig:dNdz} for the binning), {\em top right panel:} the cross correlation of the bins 1-10 and {\em bottom left panel:} the cross correlations of bins 1-4 are shown for an Euclid-like survey. For comparison, in the {\em bottom right panel} the correlation of bins 1-4 are shown also for SKA. The solid (black) line ``full" contains the correlation of density, redshift space distortion (rsd) and lensing; the dashed (blue) line ``only local"  includes density and rsd only; the short-dashed (violet) line ``local-lensing'' is the correlation of the local terms in bin $i$ with the lensing term in bin $j$, with mean redshift $\bar z_i\leq \bar z_j$.
}
\label{f:spec-cross}
\end{figure}

\subsection{Measuring the cross correlation $\langle D(z_i)\ka(z_j)\rangle$}
The situation changes drastically if we consider the cross correlation spectra.
In figure~\ref{f:spec-cross} we show them for the density, redshift space distortion and lensing terms neglecting the sub-leading Doppler and potential terms.
The ``full'' case contains the correlation of all these terms, compared to the case where we consider only ``local'' terms (density and redshift-space distortions) and the correlation of local terms in bin $i$ with lensing term in bin $j$ with mean redshift $\bar z_i\leq \bar z_ j$ (``local-lensing'' case).
Note that when fixing $\bar z_i$ and increasing $\bar z_j$, while the density and redshift space distortion terms decay, the lensing term remains substantial.
We have also checked that the Doppler and potential terms remain small for $\ell\gtrsim20$, and since larger scales are dominated by cosmic variance we neglect these effects to perform forecasts, letting a detailed study of their significance as a separate work.
The figure shows that bin auto-correlations are dominated by local terms.
However, lensing convergence dominates the cross spectra.
The correlation 1-10 is entirely given by the  density-lensing correlation (short-dashed).
For the 1-4 correlation, the density auto-correlation is still important because of the significant overlap of the wide bins 1 and 4 due to the poor photometric redshift determination.
For comparison we also show the correlation of bins 1-4 for SKA: given the spectroscopic redshift precision, this correlation is already dominated by the lensing contribution.

The largest contribution comes in particular from correlating the density transfer functions at a lower redshift bin $i$ with the integrated lensing transfer function at higher redshift bin $j$, with $i<j$.
This corresponds to a measurement of lensing of background sources by
foreground sources.
The inverse case $i>j$ gives a negligible contribution.
This shows, that for sufficiently distant bins or bins with negligible overlap like the 1-4 bin of SKA, the galaxy number counts actually measure the following quantity averaged over the respective bins with the window functions $W_{\De z}(z,z_i)$ and $W_{\De z}(z',z_j)$
\bea\label{e:Dkappa}
-b(z)(2-5s(z'))\langle D(z,\bn)\ka(z',\bn')\rangle  &=&
\frac{1}{4\pi}\sum_\ell (2\ell+1)C_\ell(z,z')P_\ell(\bn\cdot\bn')  \,.
\eea
Here $P_\ell$ is the Legendre polynomial of degree $\ell$.
All other terms contribute very little to the number count fluctuations if $r(z')-r(z)>r_c$. Here $r_c$ is a typical galaxy correlation scale of order $150h^{-1}$ Mpc, and for a mean redshift $\bar z\sim 1$, $r(z')-r(z) \gtrsim (1.76\times H_0)^{-1}(z'-z)$, this corresponds to $z'-z\gtrsim 0.09$.  For well determined redshifts, like e.g. SKA this is satisfied for all not neighboring bins in our binning. In the case of photometric redshifts the overlap of the different bins is considerable and the local terms still contribute significantly also to bins with well separated mean redshift as can be seen in the lower left panel of Fig.~\ref{f:spec-cross}.
We have verified that for the cross-correlations shown in figure~\ref{f:spec-cross}, rsd's contribute only to low multipoles $\ell<10$ dominated by cosmic variance.
However, for narrower bins, rsd can be relevant and should be added to the density term in eq.~(\ref{e:Dkappa}).

For a given (reconstructed) bias $b(z)$ and magnification bias $s(z)$, the cross correlation spectra of number counts allow us to determine the power spectrum of $\langle D(z_i,\bn)\ka(z_j,\bn')\rangle $ for sufficiently well separated redshifts, $z_j-z_i\gtrsim 0.1$.
In the next section we shall show in a simple example how bin cross-correlations can be used to constrain modifications of General Relativity on cosmological scales.

Note also the sign change between the top and the lower right panels in Fig~\ref{f:spec-cross}. This is due to the sign change in $2-5s(z')$ which happens at $s(z')=0.4$. The cross correlation spectrum of $\langle D(z,\bn)\ka(z',\bn')\rangle $ is always positive so that $C_\ell(z,z')$ given in eq.~(\ref{e:Dkappa}) is negative for $s(z')<0.4$ and positive for $s(z')>0.4$.  In our two examples, $2-5s(z)$ becomes negative for SKA in the 4th bin, while its mean remains positive for Euclid until the 8th bin.  The dependence of $s(z')$ on the limiting magnitude can actually be employed to monitor this term by varying $m_{\rm lim}$, see appendix~\ref{sec:s_bias}.

Let us finally stress that the oscillations visible in the spectra are not due to numerical precision but represent the Baryon Acoustic Oscillations visible mainly in the cross correlation spectra of the local terms.

\begin{figure}[t]
\begin{center}
\includegraphics[width=.9\textwidth]{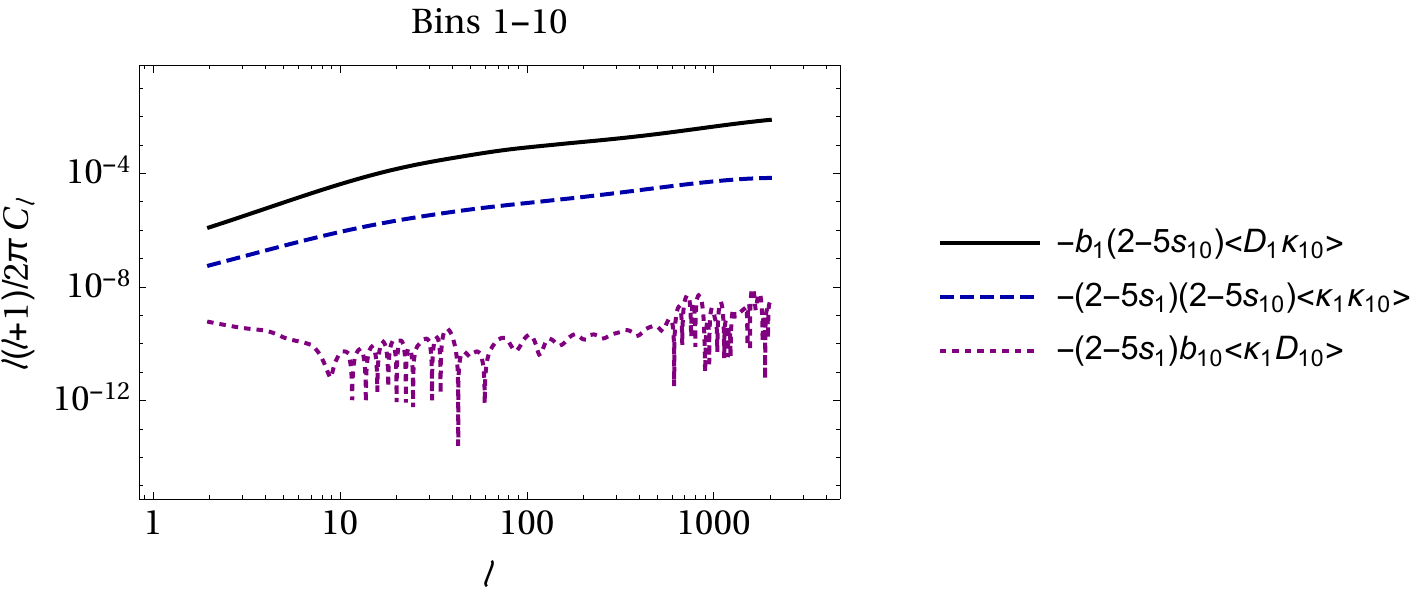}
\end{center}
\caption{The contribution from the density and convergence terms has been isolated for Euclid bins 1-10 cross-correlation. Note that since $(2-5s_i)(2-5s_j)\langle \kappa_i \kappa_j \rangle$ is negative for the chosen values of magnification bias, we plotted it with  opposite sign. Due to  the sign change of $(2-5s(z))$ at $z\approx1$, the lensing-lensing correlation is negative, while the density-lensing correlation is positive.
The numerical oscillations in $\langle \ka_1D_{10}\rangle$ are not relevant for the analysis, as the amplitude of this  term is negligible.}
\label{f:spec-cross_dk}
\end{figure}

In figure~\ref{f:spec-cross_dk} we show separately the contributions $(2-5s_i)(2-5s_j)\langle \kappa_i \kappa_j \rangle$ and $-b_i(2-5s_j)\langle D_i \kappa_j \rangle$ to the cross-correlation of the redshift bins 1 and 10.
The plot shows that the main contribution is the correlation of the density field at lower redshift with the lensing at higher redshifts, i.e., $z_i<z_j$.
This is nearly 2 orders of magnitude larger than the lensing-lensing correlation, and about 6 orders of magnitude larger than the density-lensing correlation for $z_i>z_j$.
In this last term numerical oscillations are clearly visible, but since the amplitude is negligible, they are not relevant for our Fisher matrix analysis in the next section.
The results in figure~\ref{f:spec-cross_dk} are expected, since lensing at $z_j>z_i$ is caused by all galaxies up to redshift $z_j$ hence also by the galaxies at redshift $z_i$ which are represented by $D_i$.
This explains  also why $\langle D_1 \kappa_{10} \rangle \gg \langle  \kappa_1 D_{10} \rangle$.
Since the density term is by far the most relevant one, we also expect $\langle D_1 \kappa_{10} \rangle \gg \langle \kappa_{1} \kappa_{10} \rangle $.
Again, the sign of the correlation, negative for $(2-5s_i)(2-5s_j)\langle \kappa_i \kappa_j \rangle$ and positive for $-b_i(2-5s_j)\langle D_i \kappa_j \rangle$, is also determined by the sign of the factor $(2-5s(z))$ which changes at $z\approx1$.
We have also verified that our results for the convergence spectrum are consistent with, e.g., \cite{Fosalba:2013mra}.

\begin{figure}[t!]
\begin{center}
\includegraphics[width=.45\textwidth]{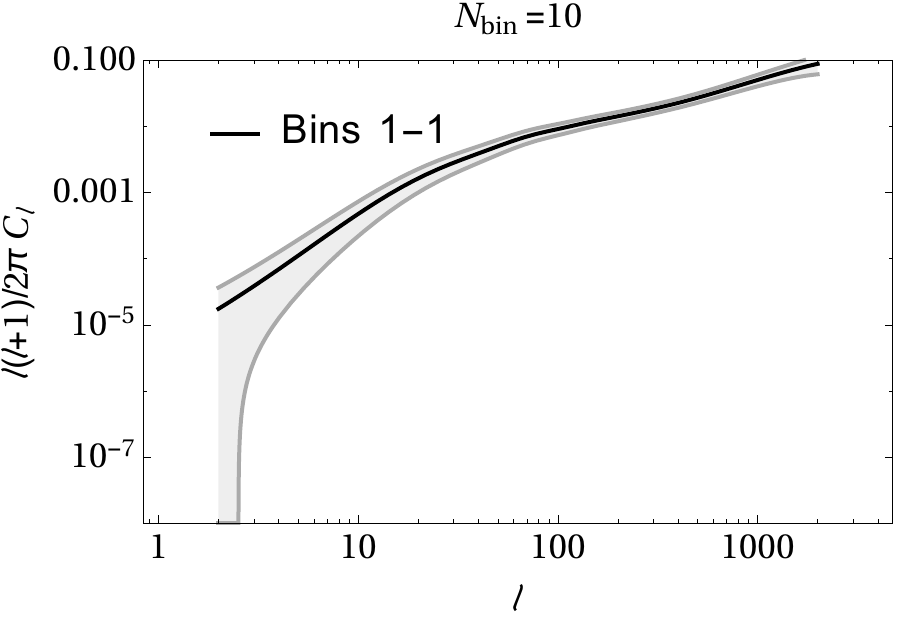}\quad
\includegraphics[width=.45\textwidth]{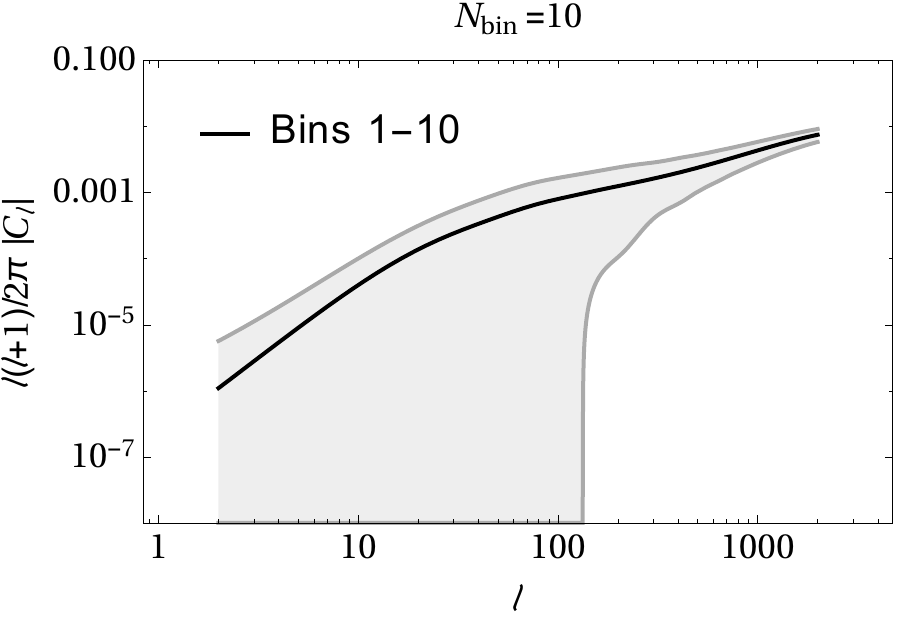}\\
\includegraphics[width=.45\textwidth]{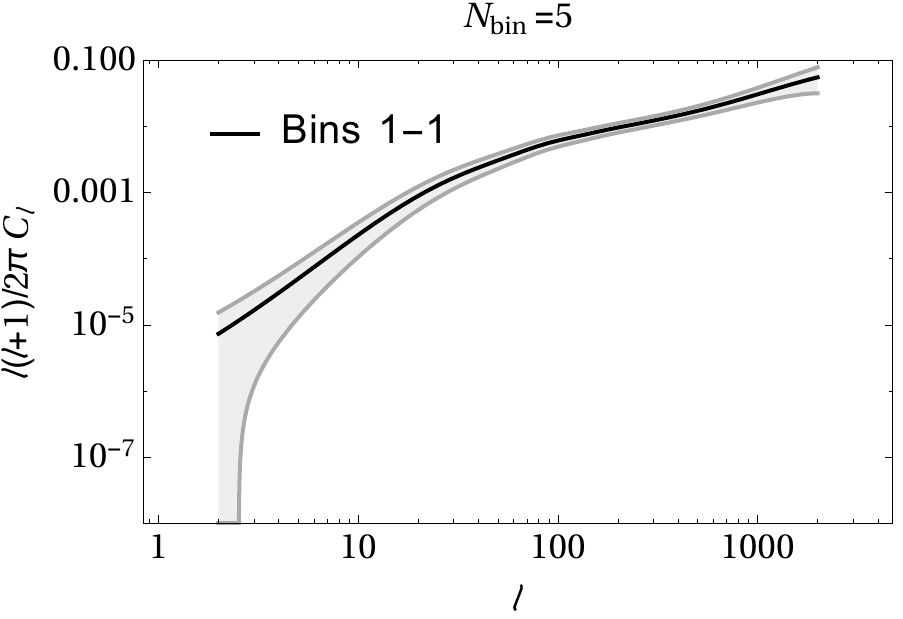}\quad
\includegraphics[width=.45\textwidth]{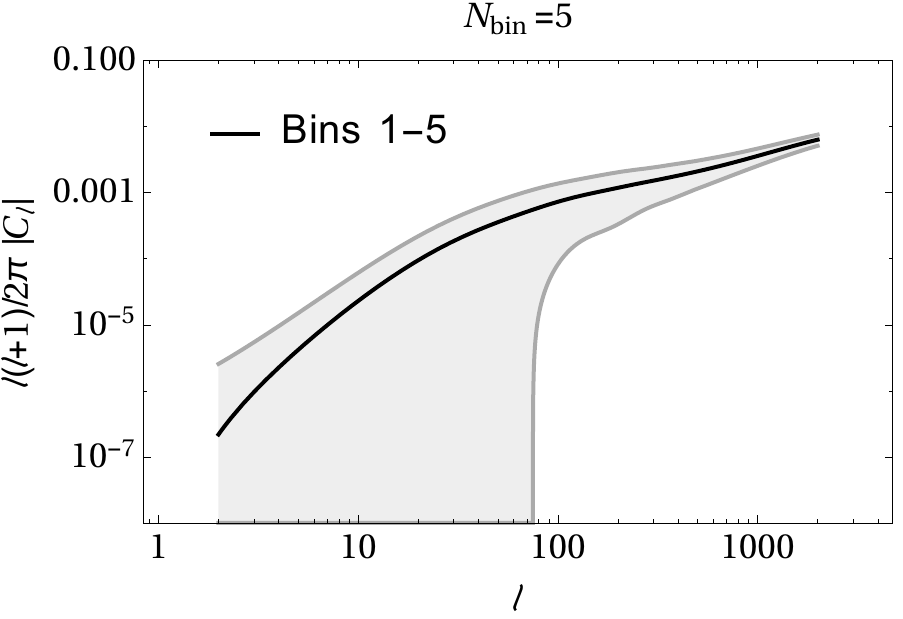}
\end{center}
\caption{
The total signal (solid line) together with the noise (gray band) are shown. The noise is dominated by cosmic variance at low multipoles and by shot-noise and theoretical errors $E_{\ell}$ at high multipoles (see appendix~\ref{sec:fisher} for details).
Upper and bottom panels consider a division into 10 and 5 equal shot-noise redshift bins, respectively.
Bin $i$-$j$ correlations are indicated inside the plot.
Thanks to lensing convergence, also cross-correlations have an important signal-to-noise for $\ell\gtrsim 100-200$.
}
\label{f:spec-noise}
\end{figure}

In figure~\ref{f:spec-noise} we also include the noise amplitudes (shaded regions) for the photometric Euclid survey with 5 and 10 Gaussian bins containing equal numbers of galaxies, see appendix~\ref{sec:surveys}.
In this plot, for illustrative purposes, we estimate the error of the signal $C_{\ell}^{ij}$ as its variance by setting $(ij)=(pq)$ in eq.~(\ref{e:Cov}):
\begin{equation}
\sigma_{ij}^2 = \frac{1}{f_{\rm sky}(2\ell+1)}\left[ C_{\ell}^{{\rm obs}, ii}C_{\ell}^{{\rm obs}, jj}+\left(C_{\ell}^{{\rm obs}, ij}\right)^2 \right] \;,
\end{equation}
where $f_{\rm sky}$ is the covered fraction of sky and $C_{\ell}^{{\rm obs},ij}$ are the observable spectra including errors, see Appendinx~\ref{sec:fisher} for more details.
Errors are dominated by cosmic variance on large scales, and by theoretical errors $E_{\ell}$ and shot-noise (see eq.~(\ref{e:Clobs})) on small scales.

For the first bin auto-correlation, both configurations show good signal-to-noise. In the case of 5 bins, the signal is slightly lower. In this case the bins are wider, hence the signal is spread out over a larger redshift range and local terms (not integrated along the line of sight) dominating auto-correlations are somewhat less important.
The cross-correlation between the first and last bins show that, despite the large redshift separation, the signal is still substantial both for 5 and 10 bins because of photometric overlap but also thanks to the contribution of lensing convergence as discussed above.
The 5 bin case has the advantage that each bin has a lower shot-noise, which  dominates the error at higher multipoles.

\begin{figure}[t!]
\begin{center}
\includegraphics[width=.75\textwidth]{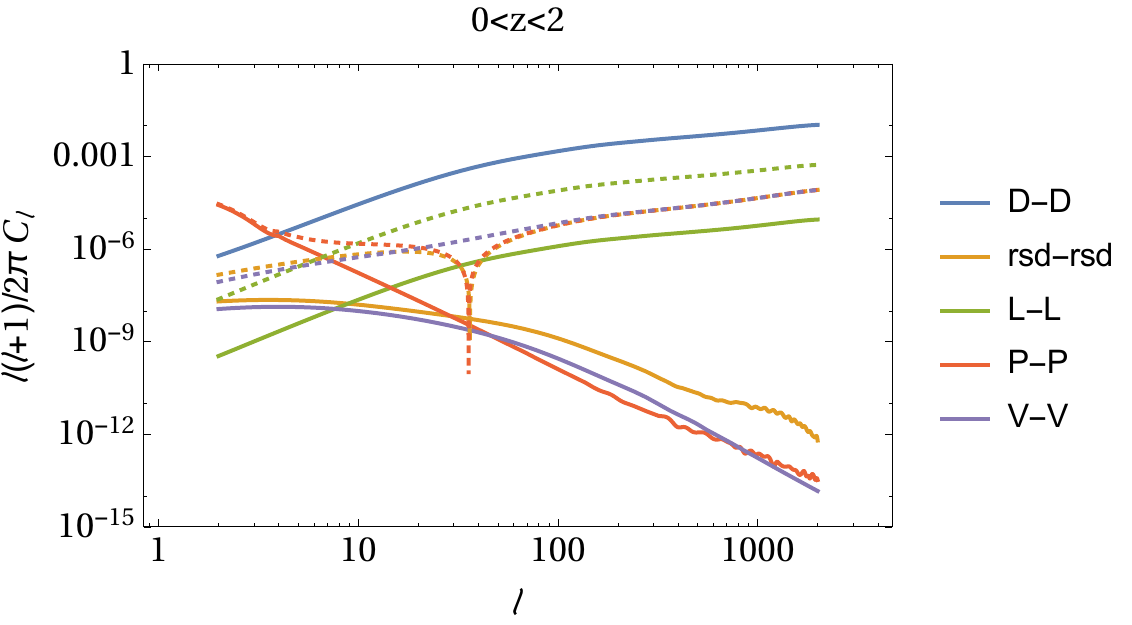}
\end{center}
\caption{The redshift integrated spectra. Compared to the 10 bin configuration, local terms are significantly reduced, while integrated terms are nearly unchanged so they become more relevant. The notation follows figure~\ref{f:spec-diag}.}
\label{f:spec-int}
\end{figure}

In figure~\ref{f:spec-int} we show the redshift integrated spectra, i.e., we choose only one bin covering the full range $0<z<2$ of the survey.
Given that the single redshift bin is now much larger than the photometric errors, we use a tophat window function.
Furthermore, we now consider the full redshift dependence of galaxy and magnification bias within the window function, unlike the cases for 5 and 10 bins where these parameters are set to their values at the mean redshift of each bin (see appendix~\ref{sec:surveys}).
Comparing the amplitude with the one in figure~\ref{f:spec-diag} shows that local terms are significantly reduced when integrating over a wide window function.
On the other hand, terms integrated along the line of side change only little, so that their relative importance increases.
Only the density term is still larger than the lensing term, while rsd and Doppler are subleading.
The potential terms now even dominate the signal at $\ell=2$ and $\ell=3$ and are the second most important term for $\ell\lesssim10$, whereas the lensing term is second for smaller scales.
As for figure~\ref{f:spec-diag}, we stress that the full signal is not given only by the auto-correlation of each term (solid lines), but we have to add all cross-correlations between different terms.
Dashed lines include  also cross-correlations of each term with the dominant density term.
Neglecting lensing would underestimated the total signal by 5-10\%.
On large scales, the potential terms become very important and can mimic a significant $f_{\bf NL}$-contribution to the bias, see \cite{Bruni:2011ta,Camera:2014bwa} for a detailed study of this effect.

\section{Lensing constraints for modified gravity from tomographic number counts}
\label{s:fm}
In this section we show that the number counts can be used to determine the lensing spectrum. For this we modify the lensing spectrum by one simple parameter,
\be
\psi(z_i,\bn) = \beta\ \psi^{\La{\rm CDM}}(z_i,\bn) \;.
\label{e:eta}
\ee
In the standard $\Lambda$CDM model we have $\beta=1$.
In the literature~\cite{Saltas:2014dha,Ade:2015rim} one often finds the so called ``slip parameter'' $\eta$ and the ``clustering parameter'' $Q$ given by
\be
\label{eq:mg_param_1}
\Phi = \eta\Psi\,, \qquad -k^2\Phi = 4\pi G a^2 Q D \,.
\ee
Here $\Phi$ and $\Psi$ are the Bardeen potentials and both $\eta$ and $Q$ can in principle depend on time and on wave number. Inserting this in eq.~(\ref{eq:lenpot})
we obtain that with these modifications, for constant values of $Q$ and $\eta$ and neglecting the  anisotropic stress from neutrinos
\be
\label{eq:mg_param_2}
\psi(z_i,\bn) = \frac{1}{2}Q(1+\eta^{-1})\psi^{\La{\rm CDM}}(z_i,\bn)  \,,
\ee
so that $\beta = \frac{1}{2}Q(1+\eta^{-1})$.
A deviation of this value from unity requires a modification of the standard $\Lambda$CDM cosmology. This can in principle be achieved, e.g., with clustering dark energy, but in most cases requires a modification of General Relativity~\cite{Saltas:2014dha}.
Generally, also the parameter $\beta$ may show a redshift and scale dependence that we neglect here for simplicity.

We recall that the lensing convergence $\kappa$ is a complementary probe to the lensing shear.
Convergence measures the trace of the deformation matrix~\cite{Bartelmann:1999yn}
\be
\mathcal{A}=\frac{\partial \bra{\theta_S,\varphi_S}}{\partial \bra{\theta_O,\varphi_O}} =
\left(
\begin{array}{cc}
1-\kappa-\gamma_1 & -\gamma_2         \\
-\gamma_2         & 1-\kappa+\gamma_1
\end{array}
\right) \;,
\ee
while shear is defined as the traceless part.
Setting
\be
\ga =\left(
\begin{array}{cc}
\gamma_1 & \gamma_2         \\
\gamma_2         & -\gamma_1
\end{array}
\right)\,.
\ee
The tensor field $\ga$ has helicity $+2$ with amplitude $\ga_1+i\ga_2$ which is also denoted by $\ga$.
Assuming a purely scalar gravitational field with lensing potential $\psi$ given in eq.~(\ref{eq:lenpot}), so that $\mathcal{A}_{ij}=\de_{ij}+\psi_{,ij}$ we have
\bea
\ka &=& -\frac{1}{2}\De\psi = -\frac{1}{4}\left(\spart\spart^* +\spart^*\spart\right)\psi\\
\ga &=& \ga_1+i\ga_2=-\frac{1}{2}\spart^2\psi \,.
\label{eq:shear}
\eea
Here $\spart$ is the spin raising and $\spart^*$ the spin lowering operator, see appendix~\ref{A:spin} and~\cite{Durrer:2008aa} for details. Expanding $\ka$ and $\psi$ in scalar spherical harmonics and $\ga$ in spin-2 spherical harmonics one finds the following relations for their power spectra (the derivation is given in appendix~\ref{A:spin}):
\be
C_\ell^\ka =\frac{\ell^2(\ell+1)^2}{4}C_\ell^\psi \;,  \qquad
\prescript{}{2}{C}_\ell^\ga =\frac{\ell(\ell+2)(\ell^2-1)}{4}C_\ell^\psi \,.   \label{eq:Clga}
\ee
Weak lensing experiments measure shear and hence $\prescript{}{2}{C}_\ell^\ga$ while the proposed lensing tomography measures $C_\ell^\ka$. The relations (\ref{eq:Clga}) therefore provide a welcome consistency check.

We determine error bars on the lensing parameter $\beta$ from a Fisher matrix analysis.
We study the dependence of our results on the number of bins for both, a Euclid-like and an SKA-like survey.
We choose the Planck 2015 results~\cite{Planck:2015xua} as our fiducial parameters around which we compute the Fisher matrix. These are the following: $\om_{\rm b}=\Omega_{\rm b}h^2=0.02225$, $\om_{\rm cdm}=\Omega_{\rm cdm}h^2=0.1198$, $n_s=0.9645$, $\ln(10^{10}A_s)=3.094$, $H_0=67.27$km/s/Mpc, $m_{\nu}=0.06$ eV and $\beta=1$, corresponding to the baryon and CDM density parameters, the primordial scalar spectral index, the amplitude of primordial curvature perturbations, the Hubble parameter, the mass of one single neutrino species (neglecting the other neutrino masses) and our lensing parameter $\beta$.
In joint constraints of a subset of parameters, we marginalize over the others.

We divide the galaxies into $N_{\rm bin}=1, 5,10$ redshift bins with the specifications described in appendix~\ref{sec:surveys}.
We assume constant galaxy and magnification bias in each bin, the values being determined by the mean redshifts.
Only for $N_{\rm bin}=1$ these parameters are fully evolved within the bin.

\begin{figure}[t!]
\begin{center}
\includegraphics[width=\textwidth]{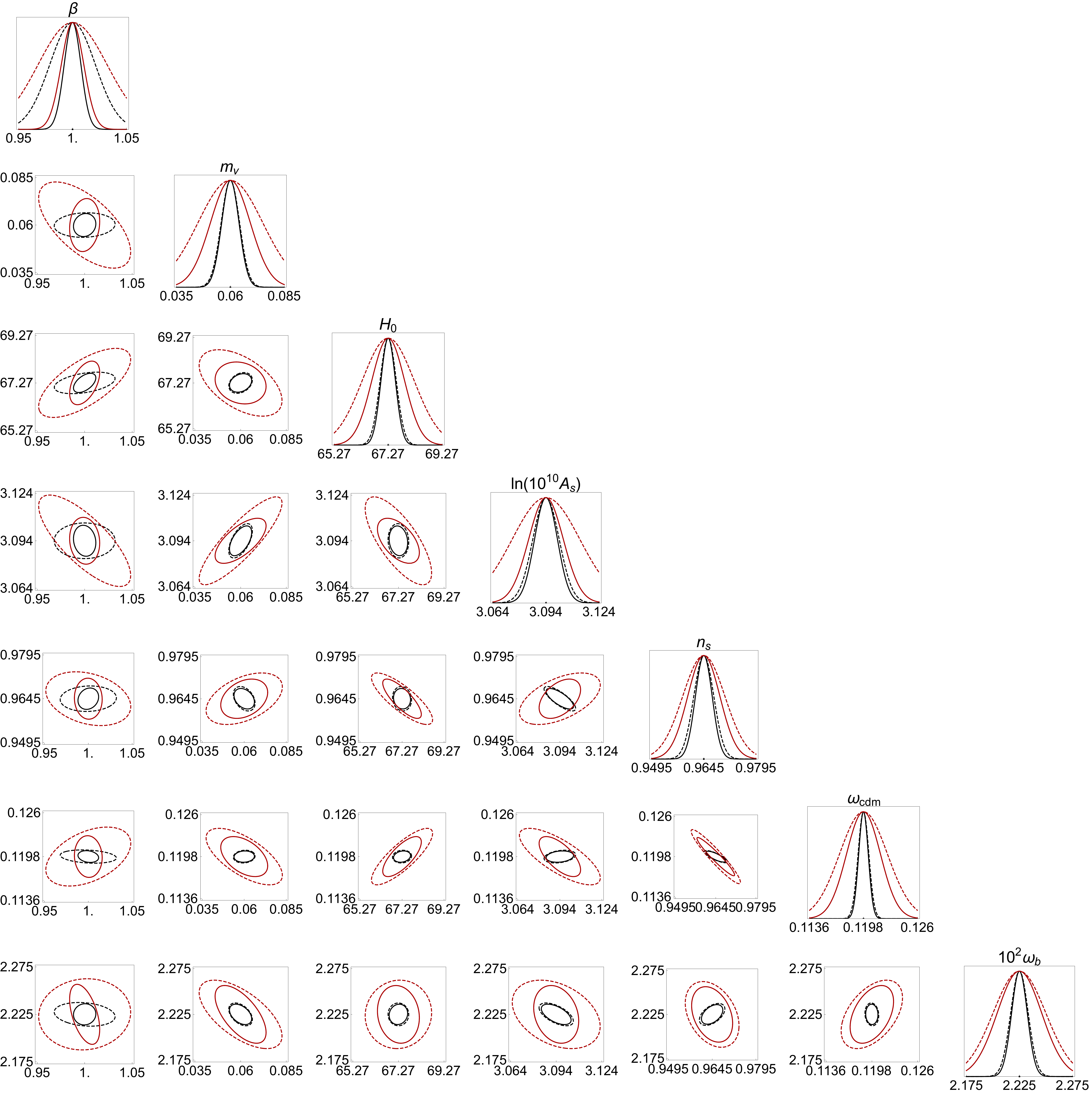}
\end{center}
\caption{
Euclid 1-$\sigma$ (68.3\%) 2d likelihood contours  and the 1d likelihood functions for 5 (red) and 10 (black) redshift bins, including all bin auto- and cross-correlations (solid) or only auto-correlations (solid). For the $\ln(10^{10}A_s)$ -- $n_s$ contour,
 the black full correlations (solid) 10-bin contour is slightly larger than the corresponding 5-bin contour.
We have checked that this is a precision issue which is not relevant for the conclusions of this work.
Note, however, that the lensing parameter $\beta$ is mostly constrained by cross-correlations. It is the only parameter for which  the 10--bin auto-correlation result (black dashed) is worse than the 5-bin cross correlation (red, solid). Also, increasing $N_{\rm bin}$ from 5 to 10 only slightly improves the precision of $\beta$ while it significantly improves the constraints on the other parameters.
}
\label{f:2d_euclid}
\end{figure}

We neglect the contribution of potential terms to the number counts, eq.~(\ref{eq:nc_pot}), relying on the fact that, as shown in figures~\ref{f:spec-diag} and \ref{f:spec-int}, they are only relevant at very large scales $\ell\lesssim20$.
These terms include effects integrated along the line of sight as lensing, and neglecting them may bring systematic errors on the determination of the lensing parameter $\beta$.
On the other hand, effects integrated along the line of sight are computationally costly and considering them, e.g., in a Markov chain Monte Carlo may be challenging.
For our purposes, actual observations can simply neglect scales $\ell\lesssim20$ which, due to cosmic variance, contribute anyway very little to the constraining power.

\subsection{Euclid forecasts}\label{s:fm_euclid}

Figure~\ref{f:2d_euclid} shows the forecasted 1-$\sigma$ contours for a Euclid-like survey.
Comparing the  cases $N_{\rm bin}=5$ (red contours) and $N_{\rm bin}=10$ (black contours), we conclude that a larger number of bins clearly improves the constraints as we expect since we add more redshift information.
Furthermore,  considering only bin auto-correlations and neglecting bin cross-correlations (dashed contours) gives of course  worse constraints than when including all correlations (solid contours).
This is particularly relevant for the lensing parameter $\beta$ since, as shown in figure~\ref{f:spec-cross}, auto-correlations are only weakly sensitive to lensing, so the constraining power comes mainly from cross-bin correlations where the density-lensing correlation is the dominant contribution.

\begin{figure}[t!]
\begin{center}
\includegraphics[width=.75\textwidth]{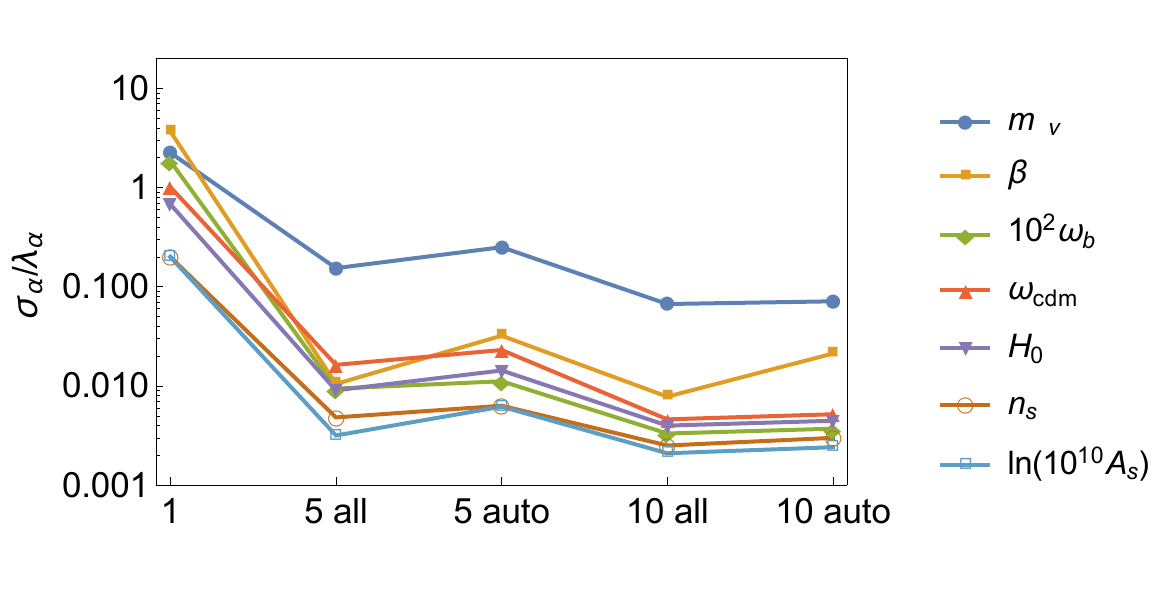}
\end{center}
\caption{
Marginalized errors (68.3\%) for Euclid, normalized to the fiducial value of a given parameter $\lambda_{\alpha}$. On the $x$-axes the numbers corresponds to $N_{\rm bin}$ and the labels indicate the situation in which all bin correlations are taken into account versus considering only auto correlations. The red line $\ln(10^{10}A_s)$ is mainly sensitive to $A_sb_0^2$, where $b_0$ is the bias at $z=0$. Since we have set $b_0=1$, marginalization over $b_0$ would substantially enhance the error of $A_s$ and, consequently, on $n_s$. Note that the precision of $\beta$ is comparable to the one of the other parameters.
}
\label{f:1d_euclid}
\end{figure}

\begin{figure}[t!]
\begin{center}
\includegraphics[width=\textwidth]{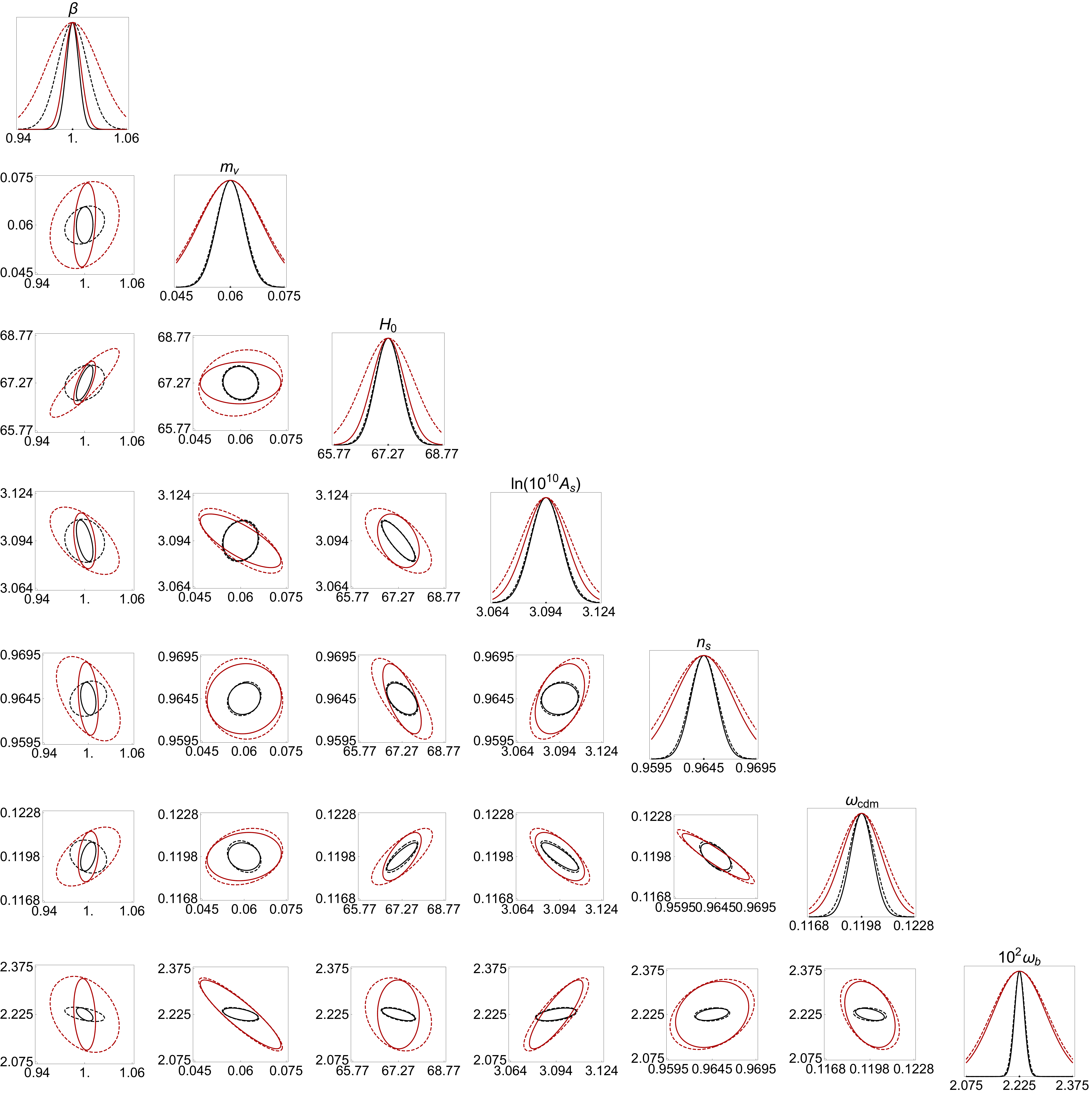}
\end{center}
\caption{
SKA 1-$\sigma$ (68.3\%) 2d likelihood contours  and the 1d likelihood functions for 5 (red) and 10 (black) redshift bins, including all bin auto- and cross-correlations (solid) or only auto-correlations (dashed). Also here some of
 the black full correlations (solid) 10-bin contours are slightly larger than the corresponding 5-bin contours due to precision. Again, for the lensing parameter $\beta$ which is mostly constrained by cross-correlations, the 10--bin auto-correlation result (black dashed) is worse than the 5-bin cross correlation (red, solid).
}
\label{f:2d_ska}
\end{figure}

In figure~\ref{f:1d_euclid} we show the 1-dimensional marginalized constraints.
In this case we also consider $N_{\rm bin}=1$. Given the large width of the bin relative to photometric errors, in this case we assume a tophat window function covering $0<z<2$.
The case $N_{\rm bin}=1$ is characterized by errors about 1-2 orders of magnitude larger than for $N_{\rm bin}=10$.
For the lensing parameter $\beta$ the error increases even by 3 orders of magnitude, confirming that most of its constraining power comes from cross-correlations of different bins.
For $N_{\rm bin}=5,10$, the forecasts suggest $\mathcal{O}(1\%)$ constraints on $\beta$.
However, because of the limitations of a Fisher analysis, rather than relying on the absolute constraint itself, we compare it to those of other standard $\Lambda$CDM parameters to conclude that we expect competitive measurement of $\beta$.
We have verified that this still holds for the more pessimistic assumption on nonlinear scales setting $\ell_{\max}=1000$ instead of 2000 (see appendix \ref{sec:fisher}), for which 1-dimensional error bars of each parameter typically increase by a factor 2-3.

As explained in appendix \ref{sec:surveys}, photometric redshift uncertainties can be modeled by Gaussian bins  (centered at the estimated true redshifts) with variance determined by the typical photometric error.
This reduces the signal, which is integrated over a broader redshift range.
In general, Gaussian redshift bins overlap and, as shown in figure~\ref{f:spec-cross}, the relative contribution from the lensing signal can be reduced for some cross-correlations, compared to the case of  non-overlapping tophat bins (e.g., the spectroscopic SKA).
On the other hand, lensing increases coherently with the width of the redshift bin while the density signal decreases.
Hence, distant correlations with small overlap are still important enough to lead to a clear improvement in the determination of $\beta$ when including all cross-correlations, as shown in figure~\ref{f:1d_euclid}.
In the next section we provide a similar study for the case of the spectroscopic SKA with tophat bins. We shall see that in both cases, including cross-correlations reduces the error in $\beta$ by a factor of about 3 for the case of 5 bins and by a factor of about 2 for 10 bins. There is no significant difference in this for Gaussian or tophat bins.

It is interesting to compare the present forecasts to those from standard galaxy clustering and weak lensing analysis.
We refer to section 1.8 of \cite{Amendola:2012ys} and to \cite{EuclidRB}.
Euclid is expected to measure the main standard cosmological parameters to percent or sub-percent level.
Our constraints on the lensing parameter $\beta$ can be compared to those on modified gravity parameters for the simplest models, as outlined in eq.~(\ref{eq:mg_param_2}).
Modified gravity constraints are expected from analysis of redshift-space distortion through the measurement of power spectra or correlation functions in Fourier space.
These observations perform better when spectroscopic redshifts are provided and constrain the matter growth factor index $\gamma_G$ defined by $\frac{d\ln G}{d\ln a}=\left(\Omega_m(a)\right)^{\gamma_G}$, where $G(a)$ is the linear growth function. The relation of $\ga_G$ to the parameters in eq.~(\ref{eq:mg_param_1}) can be found in, e.g., \cite{Hojjati:2011ix}.
The index $\gamma_G$ is  constrained by few percent for the simplest models, but the constraints weaken when marginalizing over galaxy bias, including a redshift dependence $\gamma_G=\gamma_G(z)$ or when marginalizing over several modified gravity parameters. The constraints can easily reach $\mathcal{O}(10\%)$. We expect a similar weakening of the constraints for $\beta$ when introducing, e.g., a possible redshift dependence of $\beta$.
Nevertheless, also in the simplest case, when $\beta=1$ is inconsistent with data, this already implies a deviation from the standard theory of gravity.
Therefore, this constraint can be viewed as a consistency check of GR.

Weak lensing constraints via two-point tomographic cosmic shear measurements are also very interesting for the present analysis.
The Euclid shear power spectrum is expected to be recovered to sub-percent accuracy over signal-dominated scales, giving an integrated accuracy (over all scales $\ell<\ell_{\rm max}\sim 5000$) of $10^{-7}$ \cite{EuclidRB}.
The main difficulties of shear measurements are the modeling of non-linear scales and of intrinsic alignment.
On the other hand, our measurements of the magnification $\ka$ depend significantly on the modeling of galaxy and magnification bias, but do not suffer from intrinsic alignment.
Therefore, checking the relation (\ref{eq:Clga}) is also an excellent test for systematics.
Finally, we note that while shear $\gamma$ is an observable, convergence $\kappa$ is not gauge invariant \cite{Bonvin:2011bg,Yoo:2014kpa}, but figures~\ref{f:spec-diag} and \ref{f:spec-int} prove that the terms involved in gauge transformations are subleading in our configurations.

\subsection{SKA forecasts}\label{s:fm_ska}

Figure~\ref{f:2d_ska} shows the forecasted 1-$\sigma$ contours for the SKA survey.
Both $N_{\rm bin} = 5$ (red contours) and $N_{\rm bin} = 10$ (black contours) show that including all bin correlations (solid lines) improves the constraints obtained when considering only bin auto-correlations (dashed lines).
Furthermore, in this case with very precise redshift determination, redshift bins do not overlap and are more weakly correlated compared to the photometric Euclid case.
In this case, the residual cross-correlation is nearly entirely due to integrated effects.

\begin{figure}[t!]
\begin{center}
\includegraphics[width=.75\textwidth]{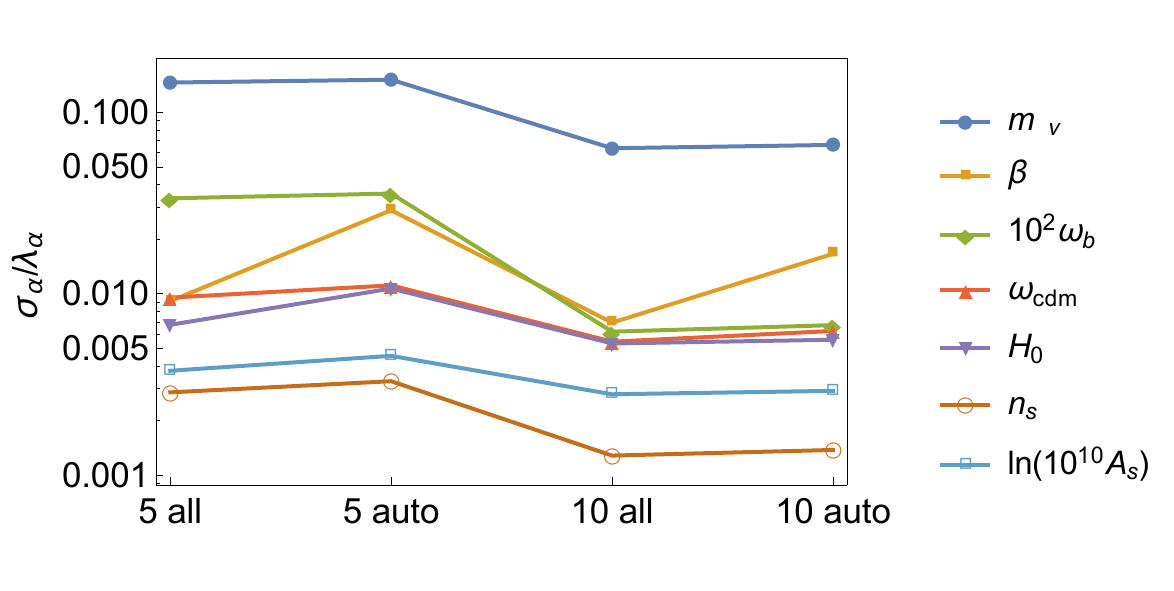}
\end{center}
\caption{
Marginalized constraints (68.3\%) for SKA, normalized to the fiducial value of a given parameter $\lambda_{\alpha}$. On the $x$-axes the numbers corresponds to $N_{\rm bin}$ and the labels indicate the situation in which all bin correlations (``all'') or only bin auto-correlations (``auto'') are included.
As for figure~(\ref{f:1d_euclid}), marginalization over $b_0$ would substantially increase errors especially on $A_s$ and, consequently, on $n_s$.
}
\label{f:1d_ska}
\end{figure}

Comparing the $N_{\rm bin} = 5,10$ contours (red lines to black lines) shows that increasing the number of bins always improves or gives similar constraints. By running parts of our simulations at higher precision, we have checked that the slight exceptions to this in some of the 2d contours are due to precision issues. The improvement with the number of bins is also shown in figure~\ref{f:1d_ska}, where we plot 1-dimensional errors. Furthermore, figures~\ref{f:2d_ska} and \ref{f:1d_ska} show that the $\beta$ parameter improves significantly when including all bin cross-correlations.

Similarly to a photometric Euclid-like survey, also the analysis for SKA suggests a constraining power on the lensing parameter similar to the errors on the other standard $\Lambda$CDM parameters.
We have also verified that, as for Euclid, this relative comparison still holds for the more pessimistic assumption on the non-linearity scales setting $\ell_{\max}=1000$ instead of 2000 (see appendix \ref{sec:fisher}), the error on each parameter again increases by a factor $\sim 2$ in this case.

Given that for $N_{\rm bin}=1$ ($0.1<z<2$) the differences between Euclid photometric and SKA spectroscopic redshifts are no longer important, we expect similar constraints from SKA as those obtained in figure~\ref{f:1d_euclid} for Euclid.

\section{Conclusions}\label{s:con}

We have investigated how the lensing potential can be constrained tomographically using the galaxy number counts statistics in future surveys.
We use the dependence of the observable number counts on the lensing convergence $\kappa$. Redshift bin auto-correlations are dominated by local terms like density and redshift-space distortions, which give small contributions to redshift bin cross-correlations.
Redshift bin cross-correlations separated by large radial distances mainly come from contributions integrated along the line of sight, which are dominated by the lensing convergence.
The number count cross correlation spectra with little overlap provide an excellent measure of the power spectrum of $\langle D(z_i,\bn)\ka(z_j,\bn')\rangle$ for $z_j>z_i$ with $\De r= r(z_j)-r(z_i)\gtrsim 150h^{-1}$Mpc, hence $\De z = z_j-z_i \gtrsim 150{\rm \ Mpc}\,h^{-1}H(\bar z)\approx 0.09$ for $\bar z\simeq1$.

Furthermore, for very deep redshift bins ($0\lesssim z \lesssim 2$), lensing  is a significant contribution also for the bin auto-correlation, confirming what has already been found in previous work~\cite{DiDio:2013sea}.
However, parameter constraints are in this case nearly 2 orders of magnitude worse compared to the subdivision into $N_{\rm bin}=5, 10$ which makes use of the redshift information.

The cross-correlations term dominating number counts for well separated redshift bins, given by the density-lensing correlation $\langle D(z_i)\kappa(z_j) \rangle$,
 corresponds to the lensing of background galaxies by foreground galaxies, and it is expected consistently with observations (e.g. \cite{Scranton:2005ci,Coupon:2013uwa}).
Lensing magnification studies usually work with two different populations, one providing the foreground and the other the background galaxies. Here we go beyond this approach and consider a fully tomographic extraction of the lensing signal as a function of redshift. Since the contribution  $\langle \ka(z_i)D(z_j) \rangle$ with $z_i< z_j$ is negligible, there is no danger of contamination. Given $D(z_i)$ from the auto-correlation spectrum, the cross-correlations term provides an excellent measure of the lensing potential via
$$\frac{\langle D(z_i)\kappa(z_j) \rangle}{\sqrt{\langle D(z_i)^2 \rangle}} \,.$$

As an application in this paper we have introduced   the parameter $\beta$ through eq.~(\ref{e:eta}).
The standard $\Lambda$CDM model expects $\beta=1$ and a deviation from this value may indicate a modification of General Relativity.
In any case, constraints on $\beta$ provide a  consistency test of the cosmological standard $\La$CDM model, which can be considered together with the measurements of other parameters like, e.g.,  the growth factor or the equation of state \cite{Amendola:2012ys} which would indicate deviations from $\La$CDM.
The proposed method provides also an independent measurement of the lensing potential, alternative to shear observations \cite{Schneider:2005ka} and it is therefore a very useful consistency check.

We have performed a Fisher matrix forecast based on a tomographic analysis with angular-redshift power spectra.
Nonlinear scales are included with a theoretical error which we model as in \cite{Audren:2012vy}, see Appendix~\ref{sec:fisher}.
We expect that both photometric Euclid-like and spectroscopic SKA-like surveys will produce constraints on the lensing parameter $\beta$ that are competitive with those on the other standard $\Lambda$CDM parameters, provided that all redshift bin cross-correlations are taken into account.
This result has been obtained by considering realistic specifications, in particular we have derived  a redshift-dependent fitting formula for the magnification bias, eq.~(\ref{eq:sz_euclid}), consistent with photometric Euclid-like surveys.

As a simple continuation of our analysis one can fit the value of $\beta$ in each redshift bin. This corresponds to a measurement of the lensing potential $\psi(z_j)$.  However, given the limitations of a Fisher forecast, a definitive estimation of errors on the lensing parameter $\beta$ or $\beta(z_i)$ should rely on a Markov chain Monte Carlo forecast. In such a forecast one would want to consider a specific experiment with its predicted likelihoods in more detail and also marginalize over the necessary nuisance parameters.
To aim at percent level constraints, our approximate treatment of non-linear scales (i.e., the rescaling of linear transfer functions by the Halofit algorithm) should be also improved.
Integrated effects other than lensing have been neglected in the present Fisher analysis, but to avoid systematic errors on the lensing parameter $\beta$ from actual observations, they should either be consistently included, or the large scales $\ell\lesssim20$ where they are significant must be excluded from the analysis (note, however,  that on this scales cosmic variance is substantial so that they do not contribute significantly to the final result).

Furthermore, it has been found that size bias, i.e., the fact that only galaxies large enough to be detected as extended sources are included in catalogs, may be even more relevant than magnification bias \cite{Schmidt:2009rh} and should be included to avoid systematic effects.

We also consider a promising future possibility to change $s(z)$ by varying  the limiting magnitude of the survey and in this way to enhance or reduce the lensing signal.

\acknowledgments
We thank Wilmar Cardona and Enea Di Dio for code comparison and useful insights; David Alonso, Daniele Bertacca, Pablo Fosalba, Martin Kunz and Alvise Raccanelli for discussions and comments; Savvas Nesseris for his publicly available data analysis codes. We acknowledge financial support by the Swiss National Science Foundation.
FM enjoyed the hospitality of the Department of Physics \& Astronomy at Johns Hopkins University, where part of this work was developed.

\newpage
\appendix
\section{Surveys specifications}

\label{sec:surveys}

\begin{figure}[t!]
\begin{center}
\includegraphics[width=.45\textwidth]{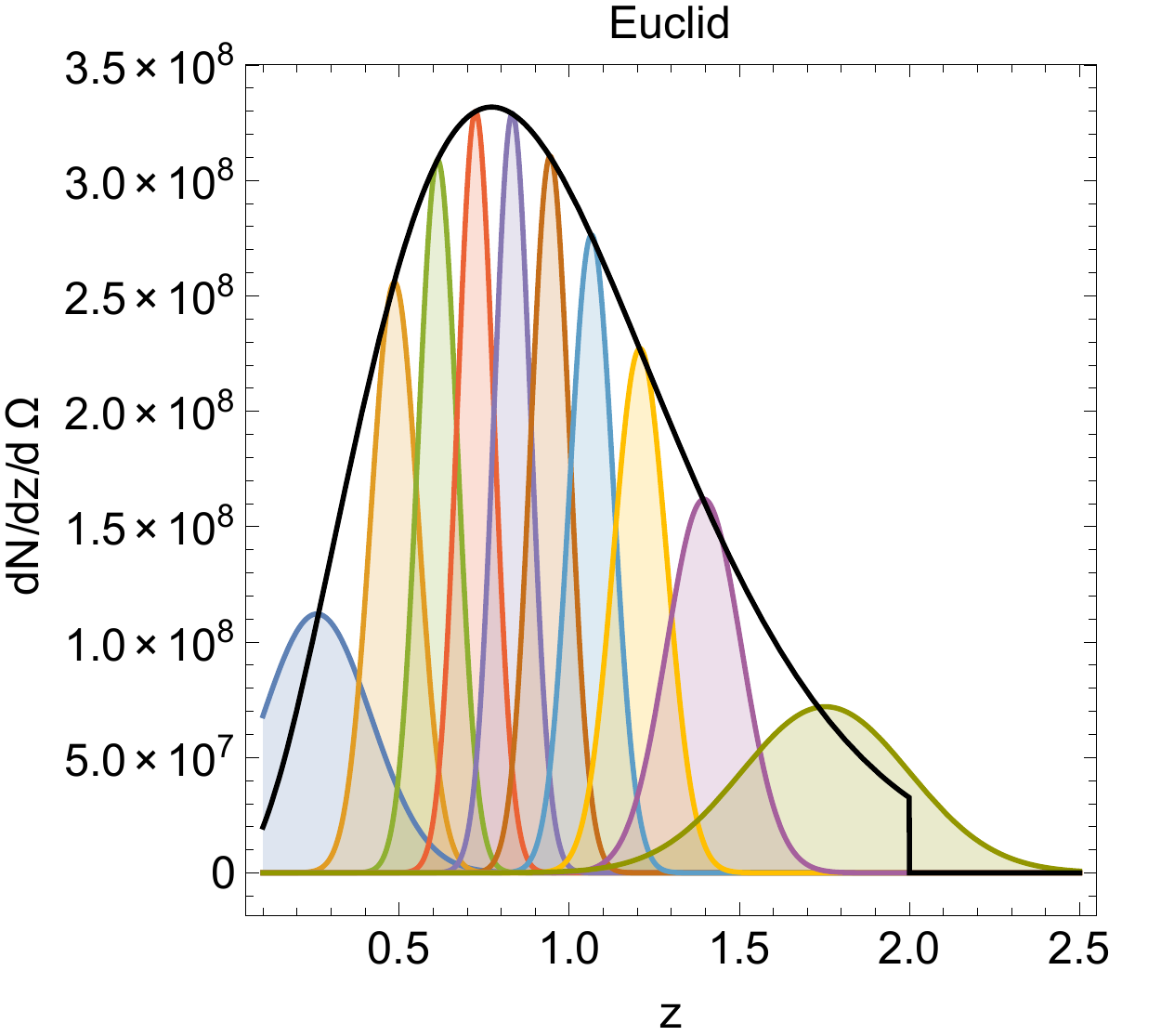}
\includegraphics[width=.45\textwidth]{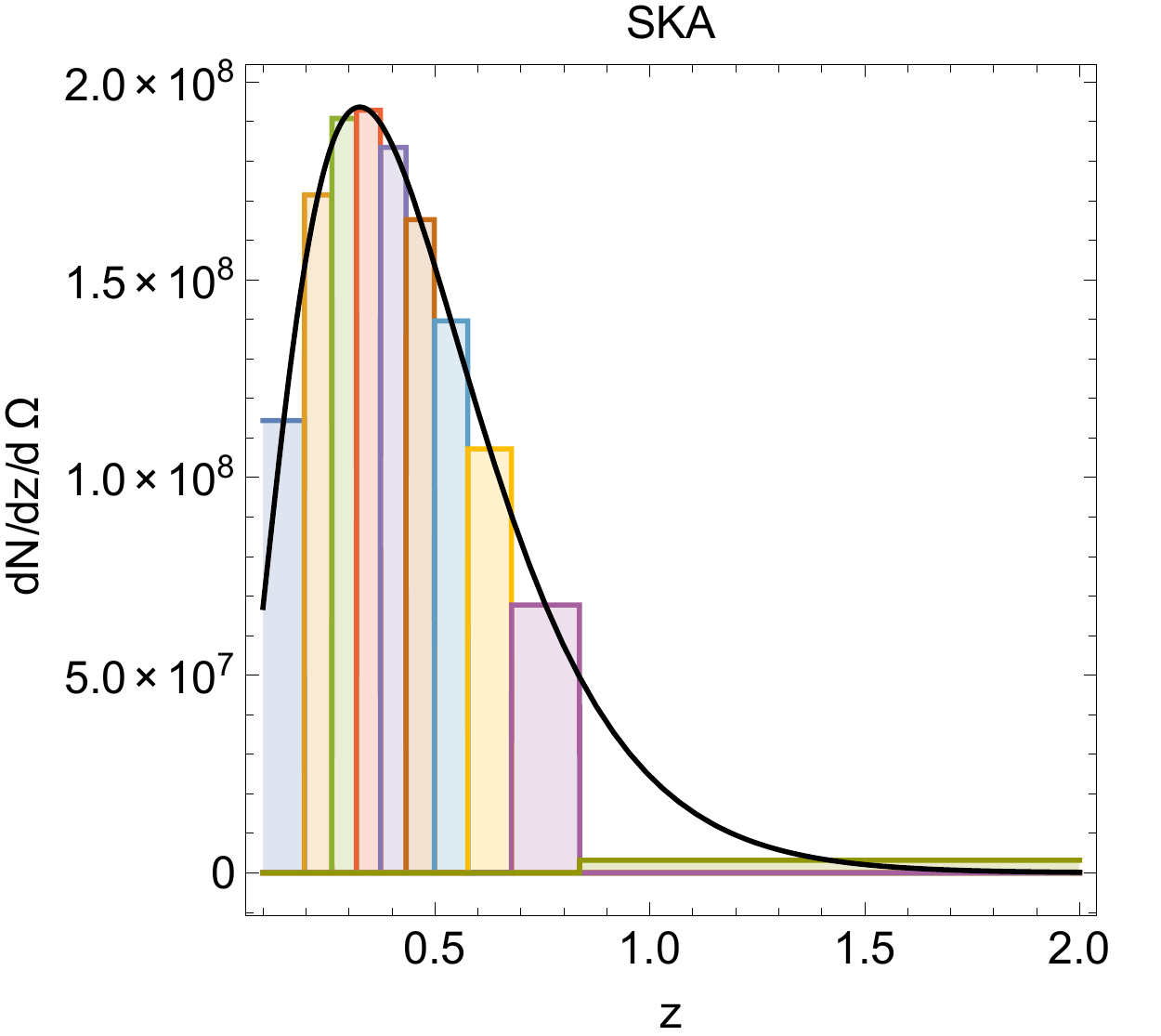}
\end{center}
\caption{Euclid (left) and SKA (right) galaxy density distribution (black line) with a division into 10 bins containing the same number of galaxies.
}
\label{fig:dNdz}
\end{figure}

\subsection{Euclid}
Following \cite{EuclidRB,Amendola:2012ys}, we consider Euclid photometric specifications and approximate the number of galaxies per redshift and per steradian, the galaxy density, the covered sky fraction, the galaxy bias and magnification bias as
\bea
&&\frac{dN}{dzd\Omega} = 3.5\times10^8 z^2 \exp\left[-\left( \frac{z}{z_0} \right)^{3/2}\right] \; \\
&&\quad \mbox{for} \quad 0<z<2.0\;, \nonumber \\
&&d=30\mbox{ arcmin}^{-2}\;,\\
&&f_{\rm sky}=0.375\;,\\
&&b(z)=b_0\sqrt{1+z}\;,\\
&&s(z)=s_0 + s_1 z + s_2 z^2 + s_3 z^3 \;. \label{eq:sz_euclid}
\eea
where $z_0=z_{\rm mean}/1.412$ and the median redshift is $z_{\rm mean}=0.9$.
We set $b_0=1$ throughout this work, but it is straight forward to include it or to marginalize over it.
The magnification bias is computed in appendix~\ref{sec:s_bias} and the coefficients are $s_0=0.1194$, $s_1=0.2122$, $s_2=-0.0671$ and $s_3=0.1031$. The full redshift and limiting magnitude dependence is given in eq.~(\ref{eq:s_z_m}).
Photometric redshift errors $\delta_z=0.05(1+z)$ allow us to choose up to $N_{\rm bin}=10$ bins such that the $i$th-bin width is $\Delta z_i \gtrsim 2\delta_z$.
Figure~\ref{fig:dNdz} shows the division into 10 bins containing the same number of galaxies, and where redshift uncertainties are taken into account by modeling the bins as Gaussian with standard deviation $\Delta z_i/2$.
For simplicity, also for $N_{\rm bin}=5$ we assume Gaussian bins with standard deviation $\Delta z_i/2$, even if in this case the bins are larger than photometric resolution (but still comparable).
For numerical convenience we set the lower redshift bound to $z=0.1$; this affects  our results by a negligible amount.
\begin{figure}[t!]
\begin{center}
\includegraphics[width=.75\textwidth]{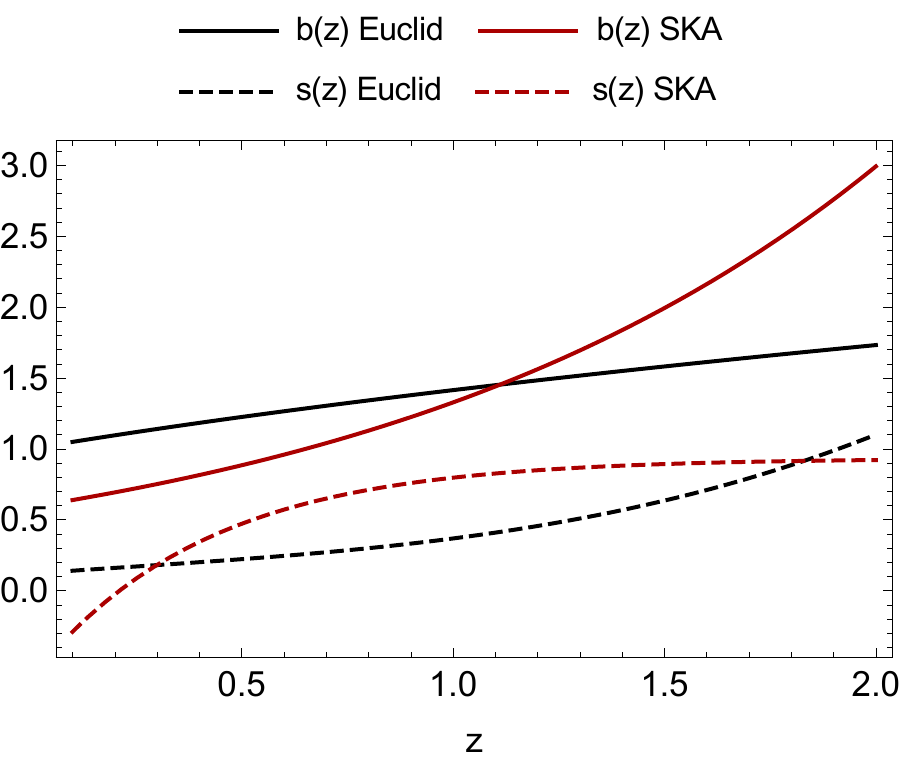}
\end{center}
\caption{Galaxy bias $b(z)$ and magnification bias $s(z)$ for Euclid and SKA. Euclid's magnification bias is computed for $m_{\rm lim}=24.5$.}
\label{fig:bs}
\end{figure}
Figure~\ref{fig:bs} shows the redshift dependence of galaxy and magnification bias. Except when specified differently, we assume constant galaxy and magnification bias in each bin, the values being determined by the mean redshifts.

For simplicity, to estimate the evolution bias $f_{\rm evo}$ (that in principle can be computed from the luminosity function), we assume to observe all the galaxies in the windows and insert $dN/dz/d\Omega$ in eq.~(\ref{eq:fevo}) to compute it.
Given the high success rate of the survey (larger than 50\%) and the fact that $f_{\rm evo}$ only appears in the subleading Doppler and potential terms, this approximation does not affect our results.

\subsection{SKA}
For SKA2 we use the specifications from \cite{Santos:2015hra,Camera:2014bwa}
\bea
&&\frac{dN}{dzd\Omega} = \left(\frac{180}{\pi}\right)^2 10^{c_1} z^{c_2}\exp\left( - c_3 z \right)\; \\
&&\quad \mbox{for} \quad 0.1<z<2.0\;, \nonumber \\
&&f_{\rm sky}=0.73\;,\\
&&b(z)= c_4 \exp\left( c_5 z \right) \;,\\
&&s(z)= c_6 + c_7 \exp\left( -c_8 z \right) \;,
\eea
where $c_1=6.7767$, $c_2=2.1757$, $c_3=6.6874$, $c_4=0.5887$, $c_5=0.8130$, $c_6=0.9329$, $c_7=-1.5621$, $c_8=2.4377$.
$dN/dz/d\Omega$ is the number of galaxies per redshift and per steradian.
The magnification bias is computed from \cite{Camera:2014bwa} considering that the specifications described above are consistent with the $5\mu$Jy sensitivity.
Given the spectroscopic redshift determination, we use tophat redshift bins.
In principle order $\sim10^2$ bins can be considered, but for comparison with the Euclid photometric case we limit ourselves to 10 bins, see figure~\ref{fig:dNdz}.
For an analysis of constraints improvement with the number of bins up to the limit set by shot-noise, see \cite{Eriksen:2015nha}.
Also in this case, except when specified differently, we assume constant galaxy and magnification bias in each bin.

\section{Euclid magnification bias}
\label{sec:s_bias}
In this appendix we derive a fitting formula for the magnification bias of the photometric Euclid survey, relying on estimates of the luminosity function.
The fit is useful to perform forecasts for Euclid-like surveys and it is not meant to be employed for precise computations.
For more involved and more accurate estimation in the optical bands see, e.g., \cite{Fosalba:2013mra}.

The Schechter luminosity function is parametrized in terms of absolute magnitudes $M$ as
\begin{equation}
\phi(M,z)dM = 0.4 \ln(10) \phi^*(z) \left(10^{0.4(M^*(z)-M)}\right)^{\alpha+1} \exp\left[ -10^{0.4(M^*(z)-M)} \right] dM \;,
\end{equation}
where $M^*(z)$ is a characteristic magnitude, $\phi^*(z)$ is a number density and $\alpha$ is a constant faint-end slope.
The Euclid photometric survey will observe in the visible and near-infrared bands in the range $550-900$nm \cite{EuclidRB}.
We use the results of \cite{Gabasch:2005bb} to model the redshift dependence of the luminosity function.
We take the $i'$ filter at $\sim770$nm as reference.
The Schechter parameters are given by (see Table 9, Case 3 of \cite{Gabasch:2005bb}):
\begin{eqnarray}
M^*(z) &=& M^*_0+a_1\ln(1+z)\;, \\
\phi^*(z) &=& \phi^*_0(1+z)^{a_2} \;, \\
\alpha &=& -1.33 \;,
\end{eqnarray}
where $a_1=-0.85$, $a_2=-0.66$, $M^*_0=-21.97$ and $\phi^*_0=0.0034\ {\rm Mpc^{-3}}$.

The cumulative number density of galaxies $\bar n(z)\equiv dN(z,M<M_{\rm lim})/dz/d\Omega$  with magnitude lower than $M_{\rm lim}$ is:
\begin{equation}
\bar n(M<M_{\rm lim}) = \int_{-\infty}^{M_{\rm lim}} \phi(M,z) dM\;.
\end{equation}
The magnification bias is defined as in eq.~(\ref{e:s_mlim}):
\begin{equation}
s(z,m_{\rm lim}) = \left.\frac{\partial\log_{10}\bar n}{\partial m}\right|_{m_{\rm lim}}\;,
\end{equation}
where $m$ is the apparent magnitude related to the absolute $M$ by
\begin{equation}
M = m - 5\log_{10}\left[ \frac{d_L(z)}{10{\rm\ Pc}} \right] - K(z)\;,
\end{equation}
where $d_L$ is the luminosity distance.
We also include $K$-corrections due to the fact that we do not observe bolometric magnitudes (integrated over all frequencies), hence a fixed observing band dims with redshift.
For a $F_{\nu}\propto \nu^{-\gamma}$ spectrum we have
\begin{equation}
K(z) = 2.5 (\gamma-1)\log_{10}(1+z)\;.
\end{equation}
Typical values in the B (ultraviolet) and K (infrared) bands are $\gamma\approx4$ and $\gamma\approx-1.5$, respectively \cite{Peacock:1999ye}.
Comparing to the $K$-corrections estimated in, e.g., \cite{Chilingarian:2010sy}, we deduce $\gamma\approx2.7$ for our $i'$ band.

Given that, at fixed redshift, derivatives w.r.t. $m$ coincide with derivatives w.r.t. $M$, we can write
\begin{equation}
s(z,m_{\rm lim}) = \left.\frac{\partial\log_{10}\bar n}{\partial M}\right|_{M_{\rm lim}}
     = \frac{1}{\ln(10)}\frac{\phi(M_{\rm lim},z)}{\bar n(M<M_{\rm lim})} \;.
\label{eq:s_z_m}
\end{equation}
Note however, that for fixed $m_{\rm lim}$, $M_{\rm lim}$ depends not only on $z$ but also on cosmological parameters via $d_L(z)$. Furthermore, we have taken into account that $d_L$ also contains fluctuations~\cite{Bonvin:2005ps,DiDio:2013bqa}.

\begin{figure}[t!]
\begin{center}
\includegraphics[width=.75\textwidth]{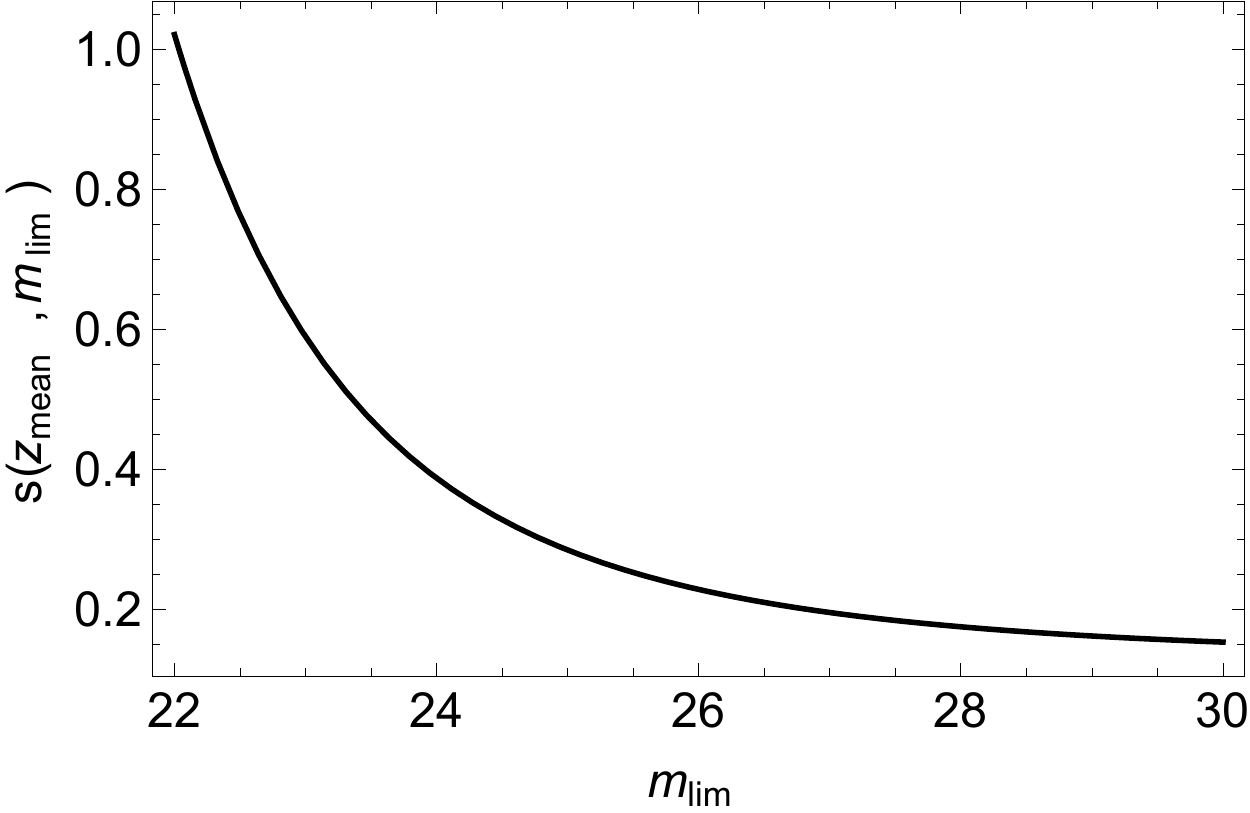}
\end{center}
\caption{Euclid magnification bias at the mean redshift $z_{\rm mean}=0.9$, as a function of the limiting magnitude.
}
\label{fig:szm_euclid}
\end{figure}

Figure~\ref{fig:szm_euclid} shows the dependence of the magnification bias on the limiting magnitude $m_{\rm lim}$ at the mean redshift $z_{\rm mean}=0.9$.
As expected, higher $m_{\rm lim}$ correspond to lower values of magnification bias.
However, for large values of $m_{\rm lim}$ magnification bias goes towards a constant value.
If the faint-end of the luminosity function would have been flat (slope $\alpha=-1$), then $s\to0$ as $m_{\rm lim}\to\infty$.

In figure~\ref{fig:bs} we also show the magnification bias for the Euclid limiting magnitude $m_{\rm lim}=24.5$ at a function of redshift. This is well approximated by the fit given in eq.~(\ref{eq:sz_euclid}).
The values $s(z=0.5)=0.22$ and $s(z=1.0)=0.37$ are consistent with the results obtained in \cite{Liu:2013yna}.
We stress that the critical value $s=2/5$ for which lensing convergence vanishes, is not an issue for our analysis thanks to the redshift dependence of $s(z)$.
One could actually  use different cuts in magnitude that would lead to an optimized $s(z)$ for our analysis. Especially, a slight increase of the limiting magnitude to e.g. $m_{\rm lim} =25$ would substantially enhance the lensing contribution.

\section{Spin weighted spherical harmonic decomposition of the lens map}
\label{A:spin}
An arbitrary tensor field on the sphere can be decomposed into its spin components. A pure spin $s$ tensor field is a symmetric completely traceless rank $s$ tensor and always has two helicity states, namely helicity $\pm s$.
These can be expanded in terms of spin weighted spherical harmonics $\prescript{}{\pm s}{Y}_{\ell m}$. The normal spherical harmonics $Y_{\ell m}$ are the   $\prescript{}{0}{Y}_{\ell m}$. See~\cite{Durrer:2008aa} for more details.
We therefore can expand the scalar functions $\ka$ and $\psi$ and the helicity $+2$ field $\ga$ as
\bea
\psi &=& \sum_{\ell m} a_{\ell m}^\psi Y_{\ell m}\\
\ka &=& \sum_{\ell m} a_{\ell m}^\ka Y_{\ell m}\\
\ga &=& \sum_{\ell m} \prescript{}{2}{a}_{\ell m}^\ga\, \prescript{}{2}{Y}_{\ell m}\, .
\eea
Applying the spin raising operator twice on $\psi$ one obtains the helicity $+2$  component of the Jacobian ${\cal A}_{ij}$ of the lens map,
\be
\ga = -\frac{1}{2} \spart^2\psi \,.
\ee
In terms of spin raising and spin lowering operators, the Laplacian is given by~\cite{Durrer:2008aa}
\be \De\psi = \frac{1}{2}\left(\spart\spart^* +\spart^*\spart\right)\psi = -2\ka \,.\ee

Expressed in terms of the polar basis $(\vartheta,\varphi)$ on the sphere the spin  raising and spin lowering operators are given~\cite{Durrer:2008aa} $\spart=-\sqrt{2}\nabla_-$ and $\spart^*=-\sqrt{2}\nabla_+$, where $\nabla_\pm$ denotes the covariant derivative in the directions of the helicity vectors $\bfe_{\pm}$ given by
$$  \bfe_{\pm}= \frac{1}{\sqrt{2}}(\dd_\vartheta \pm \frac{i}{\sin\vartheta}\dd_\varphi) \,.$$
We now use the following properties of spin weighted spherical harmonics˜\cite{Durrer:2008aa}:
\bea
\spart \left( \prescript{}{s}{Y}_{\ell m}\right) &=& \sqrt{(\ell-s)(\ell+s+1)}\, \prescript{}{s+1}{Y}_{\ell m},\\
\spart^* \left( \prescript{}{s}{Y}_{\ell m} \right) &=& -\sqrt{(\ell+s)(\ell-s+1)}\,\prescript{}{s-1}{Y}_{\ell m}.
\eea
This also implies
$$\De Y_{\ell m} = \frac{1}{2} \left(\spart\spart^* +\spart^*\spart\right)Y_{\ell m} =-\ell(\ell+1)Y_{\ell m}\, .$$
With this we find
\be
2a_{\ell m}^\ka = \ell(\ell+1)a_{\ell m}^\psi  \qquad
\prescript{}{2}{a}_{\ell m}^\ga = \frac{1}{2} \sqrt{\frac{(\ell+2)!}{(\ell-2)!}}a_{\ell m}^\psi \,.
\ee
Using the definition of the expectation values, $C_\ell^X=\left\langle|a_{\ell m}^X|^2\right\rangle$, one obtains eq.~(\ref{eq:Clga}).

\section{The Fisher matrix}
\label{sec:fisher}
We introduce the covariance matrix as in~\cite{Asorey:2012rd}
\begin{equation}
\text{Cov}_{[\ell,\ell'] [(ij), (pq)]}=\delta_{\ell,\ell'}\frac{C_\ell^{\text{obs},i p} C_\ell^{\text{obs},jq} + C_\ell^{\text{obs},i q} C_\ell^{\text{obs},jp}}{f_\text{sky} \Delta\ell \left( 2 \ell + 1 \right) },
\label{e:Cov}
\end{equation}
where $\Delta\ell$ is the bin width in multipole space.
We choose $\Delta\ell=2/f_{\rm sky}$ (rounded to the closest integer value), so that the covariance matrix is approximately block-diagonal, as empirically demonstrated in \cite{Cabre:2007rv}.
The theoretical error on non-linear scales and shot-noise are taken into account as
\begin{equation}
C_\ell^{\text{obs} ,ij}= C_\ell^{ij} + E_\ell^{ij} + \frac{\delta_{ij}}{\mathcal{N}},
\label{e:Clobs}
\end{equation}
where $E_\ell^{ij}$ is a theoretical error on power spectra, and $\mathcal{N}$ is the number of galaxy per steradian within a given redshift bin. As we determine bins with the same shot-noise, we simply have $\mathcal{N} = \frac{1}{N_{\rm bin}} \int dz \frac{dN}{d\Omega dz}$, where the integral spans over the redshift range of the survey.

To model the theoretical error $E_\ell^{ij}$, we follow the approach outlined in \cite{Audren:2012vy}.
First, we take into account non-linear corrections to power spectra by rescaling all linear transfer functions by the fit obtained by the Halofit collaboration \cite{Halofit}.
To do this, we have modified the {\sc Class} code, where this rescaling for the matter power spectrum including corrections adapted to scenarios with massive  neutrinos is  already implemented \cite{Bird:2011rb,Takahashi:2012em}.
We extrapolate the Halofit rescaling, obtained only for the density term, to all the transfer functions entering in eq.~(\ref{e:Cls}).
This gives a rough approximation for the angular-redshift power spectra $C_{\ell}^{ij}$ on non-linear scales which is good enough for our Fisher matrix forecasts on error contours.
The error power spectra $E_{\ell}^{ij}$ are obtained by taking the absolute value of power spectra computed by multiplying all the rescaled transfer functions by the square root of
\begin{equation}
\alpha(k,z) = \frac{\ln\left[1+k/k_{\rm NL}(z)\right]}{1+\ln\left[1+k/k_{\rm NL}(z)\right]} f_{\rm th} \;,
\end{equation}
where $k_{\rm NL}(z)$ is the redshift-dependent non-linear wavenumber determined by the Halofit algorithm, and $f_{\rm th}$ gives the error percentage on non-linear scales.
Even though the claimed Halofit accuracy is $\sim 5\%$, we set a more conservative $f_{\rm th}=10\%$ given that the $C_{\ell}$'s are computed by extrapolating the non-linear rescaling to all transfer functions (not only for the density).
Thus, we assume that the transfer functions are affected by a $f_{\rm th}$ error on non-linear scales $k\gtrsim k_{\rm NL}(z)$.

The Fisher matrix is then given by
\begin{equation}
\label{eq:fisher}
F_{\alpha \beta} = \sum_{\ell} \sum_{(ij) (pq)} \frac{\partial C_\ell^{ij} }{\partial \lambda_\alpha} \frac{\partial C_\ell^{pq}}{\partial \lambda_\beta} \left[\widehat{\text{Cov}}^{-1}\right]_{\ell, (ij), (pq)},
\end{equation}
where the $\lambda_\alpha$ indicate cosmological parameters.
We sum all multipoles up to $\ell_{\rm max}=2000$, as errors on non-linear scales are taken into account via $E_{\ell}^{ij}$.
The second sum is over the matrix indices $(ij)$ with $i\le j$ and $(pq)$ with $p\le q$ which run from 1 to
 $N_\text{bin}$ when considering all bin auto- and cross-correlations.
The hat on the covariance matrix indicates that it should be first reduced to the   correlations $(ij), (pq)$ which are considered in the sum and then inverted.
E.g., when neglecting the signal from bin cross-correlations only $i=j$ and $p=q$ are kept in the sum, and the Cov matrix of dimension $\left[N_{\rm bin}(N_{\rm bin}+1)/2\right] \times \left[N_{\rm bin}(N_{\rm bin}+1)/2\right]$  (for fixed $\ell$) is first reduced to a matrix $\widehat{\text{Cov}}$ of dimension $N_{\rm bin} \times N_{\rm bin}$ and then inverted.

When constraining a subset of cosmological parameters we marginalize over the remaining ones.
Hence, confidence levels are determined by \cite{Durrer:2008aa}
\begin{equation}
\Delta\chi^2 = \sum_{\alpha,\beta} \left( \lambda_{\alpha}-\overline{\lambda}_{\alpha} \right) \left[\left(\widehat{F^{-1}}\right)^{-1}\right]_{\alpha \beta} \left( \lambda_{\beta}-\overline{\lambda}_{\beta} \right) \;,
\label{e:chi2}
\end{equation}
where the hat again indicates the sub-matrix of $F^{-1}$ obtained by considering only lines and columns corresponding to the sub-set of parameters under considerations.
Fiducial values are indicated by an overbar $\overline{\lambda}_{\alpha}$.
The sum runs over the parameters being constrained.
For 1-dimensional and 2-dimensional contours, 1-$\sigma$ ($68.3\%$) confidence levels are determined by $\Delta\chi^2=1$ and 2.30, respectively.
Note that the marginalized 1-dimensional error on the parameter $\lambda_{\alpha}$ is simply given by $\sigma_{\alpha}=\sqrt{\left(F^{-1}\right)_{\alpha\alpha}}$, while we compute the 2-dimensional contours shown in Figs.~\ref{f:2d_euclid} and \ref{f:2d_ska} by solving eq.~(\ref{e:chi2}).


Finally, numerical derivatives in eq.~(\ref{eq:fisher}) are computed via the five-point stencil algorithm, see Table 25.2 of \cite{Abramowitz:1972}:
\begin{equation}
\frac{\partial C_\ell^{ij} }{\partial \lambda_\alpha} \approx
\frac{-C_\ell^{ij}(\lambda_\alpha+2h_\alpha)+8C_\ell^{ij}(\lambda_\alpha+h_\alpha)-8C_\ell^{ij}(\lambda_\alpha-h_\alpha)+C_\ell^{ij}(\lambda_\alpha-2h_\alpha)}{12h} \;.
\end{equation}
Here we indicate the dependence of the spectra $C_\ell^{ij}$ on cosmological parameters $\lambda_\alpha$.
Besides being affected by small numerical errors scaling with steps and fifth derivatives as $\frac{h_\alpha^4}{30} C_\ell^{ij\;(5)}(c)$, where $c\in\left[ \lambda_\alpha-2h_\alpha , \lambda_\alpha+2h_\alpha \right]$, the algorithm is less sensitive to the $h_\alpha$ step choice than the 2-point derivatives.
We choose $h_\alpha$ of the same order as the 1-dimensional errors (in this case obtained without marginalization) and we have verified that these are stable when further  reducing the steps.

\bibliography{lens-refs}
\bibliographystyle{JHEP}
\end{document}